\documentclass[final,authoryear,3p,times]{elsarticle}
\usepackage{amsmath,amsfonts,amssymb} \textwidth 16cm \textheight 21cm
\usepackage{graphicx,amsmath} \usepackage{epstopdf}
\usepackage{relsize} \usepackage{color} \usepackage{float} %\date{}
\usepackage{MnSymbol} \usepackage{subfigure}
\usepackage[toc,page]{appendix}
\def\micron{\:\mu\mbox{m}} \let\eps\varepsilon %\usepackage{mdframed}

\newcommand{\bea}{\begin{eqnarray}} \newcommand{\eea}{\end{eqnarray}}

 \newcommand{\matr}[1]{\mathbf{#1}}

\journal{Journal of the Mechanics and Physics of Solids}

\begin{document}

\begin{frontmatter}

\title{Learning crystal plasticity using digital image correlation: Examples from discrete dislocation dynamics}
\author[WVU,JHU]{Stefanos Papanikolaou}
\author[WVU]{Michail Tzimas}
\author[WVU,JHU]{Hengxu Song}
\author[NISTMML]{Andrew C.E. Reid}
\author[NISTITL]{Stephen A. Langer}

\address[WVU]{The West Virginia University, Department of Mechanical $\&$ Aerospace Engineering, Morgantown, WV, 26505}
\address[JHU]{The Johns Hopking University,Department of Mechanical Engineering,3400 N Charles St, Baltimore, Maryland, 21218}
\address[NISTMML]{Material Measurement Laboratory, National Institute of Standards and Technology, Gaithersburg, MD 20899}
\address[NISTITL]{Information Technology Laboratory, National Institute of Standards and Technology, Gaithersburg, MD 20899}

\begin{abstract}

Digital image correlation (DIC) is a well-established, non-invasive technique for tracking and quantifying the deformation of mechanical samples under strain. While it provides an obvious way to observe incremental and aggregate displacement information, it seems likely that DIC data sets, which after all reflect the spatially-resolved response of a microstructure to loads, contain much richer information than has generally been extracted from them. In this paper, we demonstrate a machine-learning approach to quantifying the prior deformation history of a crystalline sample based on its response to a {\em subsequent} DIC test. This prior deformation history is encoded in the microstructure through the inhomogeneity of the dislocation microstructure, and in the spatial correlations of the dislocation patterns, which mediate the system's response to the DIC test load. Our domain consists of deformed crystalline thin films generated by a discrete dislocation plasticity simulation. We explore the range of applicability of machine learning (ML) for typical experimental protocols, and as a function of possible size effects and stochasticity. Plasticity size effects may directly influence the data, rendering unsupervised techniques unable to distinguish different plasticity regimes.

\end{abstract}

\end{frontmatter}

\section{Introduction}

The prior deformation history and/or material processing of mechanical components may dramatically influence their mechanical properties without nominally altering their chemical composition~\citep{frazier2014}. Given a mechanical component design, it is mandatory to identify whether the mechanical properties of the component are in accord with its nominal properties. In crystals, a major component of property sensitivity is plastic deformation. How can we inexpensively and swiftly distinguish whether a crystalline sample is pristine or if it has been heavily deformed in the past? Scanning electron microscopy (SEM), X-ray powder diffraction, and Transmission Electron Microscopy (TEM), are insightful but relatively cumbersome ways of answering this question. Moreover, they may be consistently successful for elasticity mapping or for  molecular microstructures with the same crystal type but different interatomic potentials~\citep{kalidindi2015} but it is not clear what to expect when plastic inhomogeneity and strain localization are present, as in ductile metals and alloys. In contrast, nanoindentation or digital image correlation (DIC)~\citep{anuta1970,keating1975} are both inexpensive, swift and applicable to a vast range of material classes. In this paper, we focus on DIC: Is it possible to devise a technique for generic quantification of prior plastic deformation in crystals by using DIC? What would be the basic requirements of such an endeavour?

For crystal plasticity, the principal signature of plastic deformation has been strain localization, typically in the form of shear bands~\citep{bigoni1991,asaro-book}. However, this is only a fragmentary and incomplete measurement, and for a complete characterization of plasticity, full strain correlation signatures should be taken into account. Moreover, computing the full strain correlation signatures becomes mandatory at the microscale, where size effects and stochastic deformation can be a defining factor for plastic deformation. A particularly popular method for tracking local strain fields is digital image correlation (DIC)~\citep{anuta1970,keating1975}.  DIC is crucially dependent on optical resolution standards as well as frame rate, rendering difficult the direct observation of strain localization in crystal plasticity. Experimental studies have shown that high-resolution atomic force microscopy (AFM) has reached values less than a nm, to calculate step heights of a grain boundary in the microstructure~\citep{chen2013}. In contrast, studies on fatigue crack growth have shown that obtaining multiple optical microscope images (like DIC does) of cracks is possible and inexpensive, with resolution of 10 nm and more~\citep{chen2011,chendiss,chen2012,carroll2013,khademi2015,khademi2016a,khademi2016b,khademi2017}, through coupling with transmission electron microscopy (TEM) or electron backscatter diffraction (EBSD)~\citep{chen2012,chen2011}. In this paper, we emulate DIC using discrete dislocation dynamics (DDD) simulations, in order to quantify and classify the statistical correlations of various plastic deformation stages.

Strain correlation patterns in a microstructure may be investigated through the material knowledge system (MKS) framework~\citep{fast2011,fastkalidindi}, which can efficiently quantify generic correlations. MKS uses statistical approaches to create processing-structure-property relationships and has the potential to bridge multiple length scales using localization and homogenization linkages. MKS is made more practical and flexible by the Python Materials Knowledge System (PyMKS)~\citep{wheeler2014} framework, which is an object-oriented set of tools and examples, written in Python, that provide high-level access to the MKS frameworks for rapid creation and analysis of structure property-processing relationships. We will be using the MKS and PyMKS frameworks along with statistical tools from scikit-learn~\citep{pedregosa2011} for efficiently classifying correlations.

Mitchell defines machine learning (ML) as a natural outgrowth of the intersection of Computer Science and Statistics~\citep{mitchell2006}. ML has rapidly evolved in the last decade to include almost all scientific communities~\citep{russell1995,friedman1998,werning2010,james2013}. From the most simple Gaussian representation of a data set~\citep{rasmussen} to complex data mining applications and evaluation, ML has made major inroads within materials science and holds considerable promise for materials research and discovery. Some examples of successful applications of ML within materials research include accelerated and accurate predictions (using past historical data) of phase diagrams, crystal structures, and material properties~\citep{mueller2016}.  

\begin{figure}[H] \centering
\includegraphics[width=0.89
\textwidth]{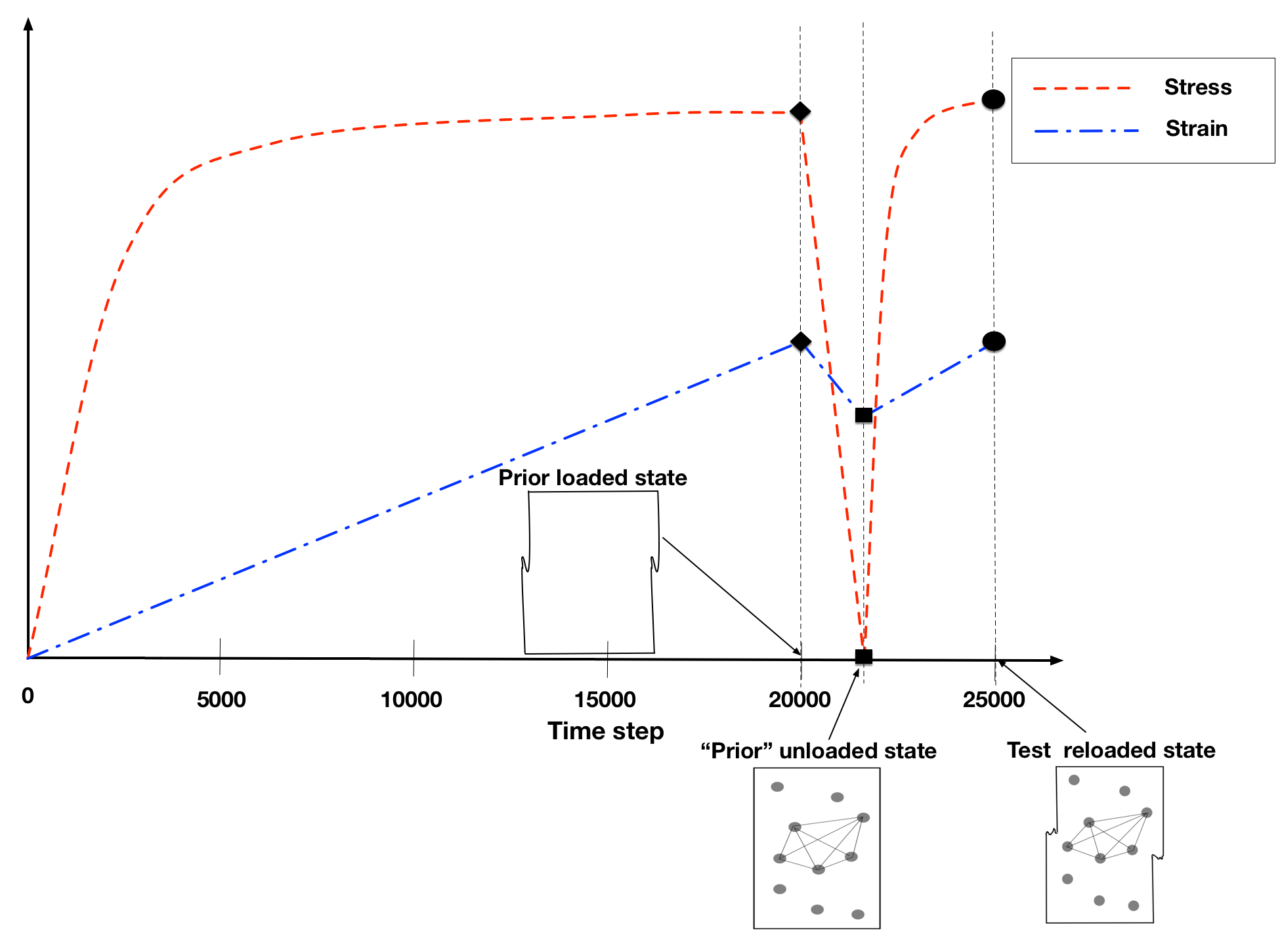}
\caption{\textbf{Schematic of the DIC protocol in modeling and sequence of sample loading:} 
A schematic of the sequence of events when loading, unloading and
reloading a sample is shown, with the corresponding stress and strain
graphs vs the time step of simulation.  A sample is obtained from
2D-DDD simulation and it's loaded to a specific strain value -- Prior
loaded state (found at specific time steps). Then, the sample is
unloaded to zero stress and the remaining plastic strain can be
calculated -- Prior unloaded states. Finally, the sample is reloaded
to a DIC-testing strain -- Test reloaded state. The DIC process begins when the sample is at the prior unloaded state. Even though a sample has been plastically deformed (at the prior loaded state), the samples obtained for DIC experiments can be polished, thus the surface of a sample is not able to provide information about deformation (the prior unloaded state sample can be seen in the figure as having smooth surface). Randomly placed DIC-tracking nanoparticles that are detected optically and contribute to correlation statistics are applied to the sample. Then, as the sample is reloaded, the permament deformation can be seen via DIC, since there are changes in the distances between tracked nanoparticles.}
\label{fig:schematic1}
\end{figure}

In our work, we use ML for quantifying plastic deformation in crystals. A simple example of crystal plasticity arises during uniaxial compression of single crystals. In this context, the simplest realistic case is a thin film, which can be modeled by simple 2D-edge discrete dislocation dynamics (2D-DDD) simulations~\citep{vandergiessen1995}. Such 2D simulations have been able to explain experimentally observed aspects of size dependent plasticity in thin films~\citep{nicola2006,deshpande2012,vlassak2014,deshpande2013,shishvan2010}. The benefit of 2D-DDD for thin films is that it allows for efficient numerical evaluation of inter-dislocation correlations. Nevertheless, the methods developed here are directly applicable to more complex numerical approaches such as 3D-DDD~\citep{ghoniem2006} or continuous crystal plasticity models~\citep{asarolubarda}, or even actual DIC experiments. Furthermore, thin films have demonstrated a rich phenomenology, since they display strong size effects and stochastic plastic fluctuations as the sample thickness decreases~\citep{papanikolaou2017,vandergiessen2006}. In this paper, we explore whether such size effects and stochastic fluctuations influence the applicability of ML methods.  In this paper, we build an approach for explicit statistical analysis of plastically deformed microstructures by mimicking the DIC protocol (Fig.~\ref{fig:schematic1}). We simulate thin film compression with 2D discrete dislocation dynamics and then we use correlation algorithms as a statistical approach, analogous to detecting digital image correlations in thin films, as a way to infer the existing deformation history from total strain profiles. Such strain profiles may originate in non-invasive mechanical testing (such as DIC or Digital Strain Imaging), and we assume they do. Nevertheless, the proposed method is applicable to any strain profile, however it is obtained. ML protocols help build our statistics model for recognition of differently deformed microstructures. Principal component analysis (PCA) is used along with 2-point correlation statistics~\citep{niezqoda2008} to help us recognize and find possible correlations of slip bands formed due to stress, and then a clustering algorithm is used to characterize these possible correlations in an efficient way. 

The remainder of this paper is organized as follows: In section 2 we describe our 2D-DDD model for obtaining mock data samples and we also describe our approach for quantifying strain correlation patterns. In section 3 we discuss the ML approaches we use for processing our data samples, and present our results based on various parameters. In section 4 we present a summary of our work and the conclusions drawn.

\begin{figure}[H] \centering
\includegraphics[width=0.5\textwidth]{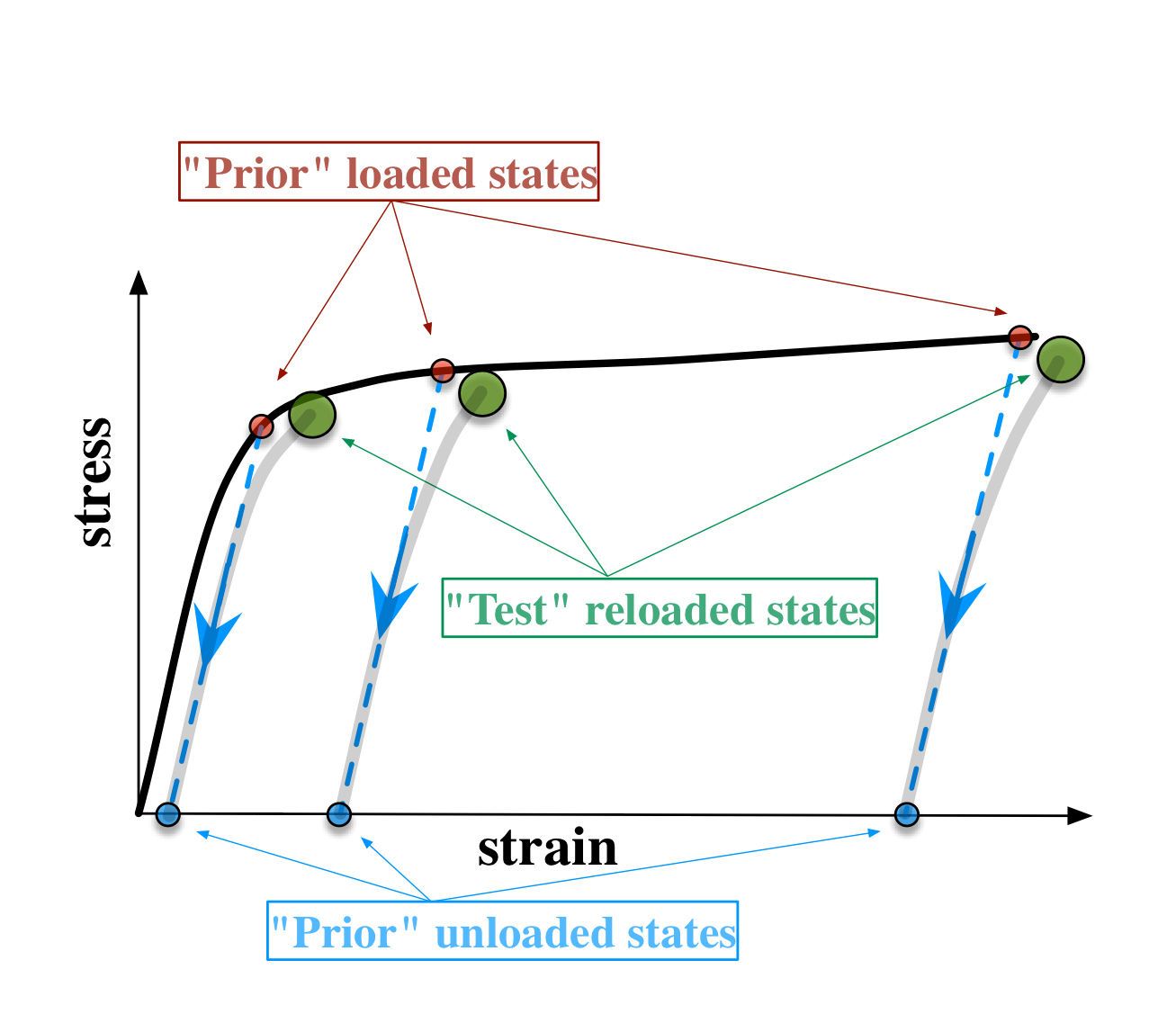}
\caption{\textbf{DIC for various loading histories:}
A material has an assumed prior deformation history (prior loaded states). We unload the sample and obtain the prior unloaded states. How does the result of DIC, which characterizes the test reloaded states, reflect the prior history?}
\label{fig:schematic2}
\end{figure}

\section{The model}
\label{sec:basic}
\subsection{Discrete dislocation dynamics}

\begin{figure}[H] \centering
\includegraphics[width=\textwidth]{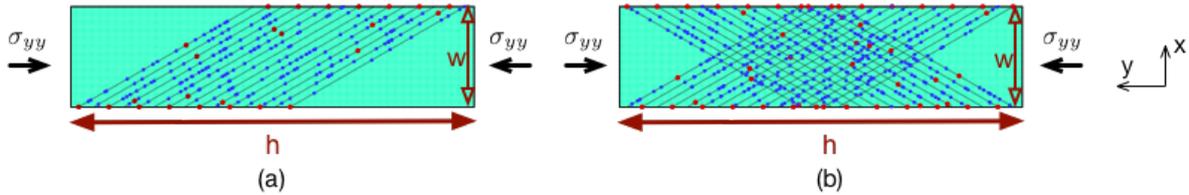}
\caption{\textbf{The 2D discrete dislocation plasticity model of uniaxial compression of thin films:} Slip planes (lines) span the sample, equally spaced at $d=10b$, but planes close to corners are deactivated to maintain a smooth loading boundary. Surface and bulk dislocation sources are present (disks) and forest obstacles are spread homogeneously across the active slip planes. Initially the sample is free of stress and mobile dislocations (a) Single slip system (b) Double slip system}
\label{fig:schematic}
\end{figure}

The geometry of the model problem is shown in Fig.~\ref{fig:schematic} for single slip system (a) and double slip systems (b). In this work, our primary focus will be on the double slip system samples, and we will compare the performance with single slip in Section 3.3. Samples are modeled~\citep{papanikolaou2017} by a rectangular profile of width $w$ and aspect ratio $\alpha$ ($\alpha=h/w$).

Plastic flow occurs by the nucleation and glide of edge dislocations, on single or double slip systems. We study sample widths ranging in powers of 2 from $w=0.125$ (or $w_0$) to $2 \micron$ with $\alpha=h/w=4$. The top and bottom edges ($x=0,w$) are traction free, allowing dislocations to exit the sample. Loading is taken to be ideally strain-controlled, by prescribing the $y$-displacement at the lateral edges ($y=0,h$). The applied strain rate (for both loading and unloading regimes), $\dot{h}/h=10^4~\text{s}^{-1}$, is held constant across all our simulations, similar to experimental practice. Plastic deformation of the crystalline samples is described using the discrete dislocation framework for small strains~\citep{vandergiessen1995}. Each dislocation is treated as a singularity in a linear elastic background solid with Young's modulus $E$ and Poisson ratio $\nu$, whose analytic solution is known at any position. 

\begin{figure}[H] \centering
\includegraphics[width=\textwidth]{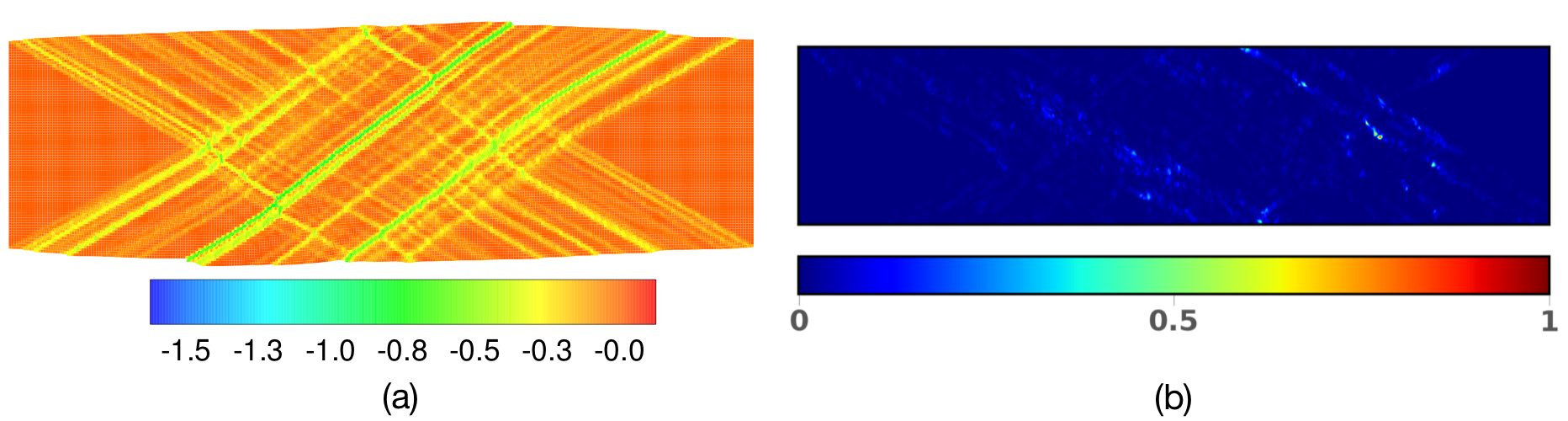}
\caption{\textbf{Strain profiles captured from the 2D-DDD simulation at the "prior" loaded and "test" reloaded states -- $\boldsymbol{w=2}~\boldsymbol{\mu \rm{m}}$ -- Double slip system:} A sample is loaded to 10 $\%$ strain and the resultant strain profile is shown. Then, the sample is unloaded to zero stress and is reloaded to DIC-testing strain of 0.1~$\%$. (a) Strain profile for the prior loaded sample at 10 $\%$ strain. Plastic steps are allowed to emerge on the film surface~\citep{papanikolaou2017}. The strain map is unitless. (b) A strain profile at the test-reloaded state is shown, after subtracting the residual plastic deformation at the prior-unload state. We consider such strain profiles as analogous to typical DIC experimental strain profiles. For higher strain localization we have higher values in the colormap (light blue to red), while for little to no strain localization we have smaller values in the colormap (darker blue). Since we subtract the remaining plastic strain at the prior unloaded states, all the previous localizations have been removed from the strain profile, thus most of the strain profile has a dark blue color (0 values in colormap). The strain map is unitless.}

\label{fig:strain_profiles}
\end{figure}

% Image fields are obtained by solving a linear elastic boundary value problem using finite elements with the boundary conditions changing as the dislocation structure evolves under the application of mechanical load.
Available slip planes are separated by $10$ Burgers vectors and are oriented at $\pm 30^\circ$ from the loading direction (Fig.~\ref{fig:schematic}).  At the beginning of the calculation, the crystal is stress free and there are no mobile dislocations.  Bulk sources are randomly distributed over slip planes, and their strength is selected randomly from a Gaussian distribution with mean value $\bar{\tau}_{\rm nuc} = 50$ MPa and 10 $\%$ standard deviation. We only consider glide of dislocations, neglecting the possibility of climb. The motion of dislocations is determined by the component of the Peach-Koehler force in the slip direction. Each sample contains a random distribution of forest dislocation obstacles and surface and bulk dislocation sources.  Once nucleated, dislocations can either exit the sample through the traction-free sides, annihilate with a dislocation of opposite sign when their mutual distance is less than $6b$, or become pinned at an
obstacle. Our simple obstacle model is that a dislocation stays pinned until its Peach-Koehler force exceeds the obstacle-dependent value $\tau_{\rm obs}b$. The strength of the obstacles $\tau_{\rm obs}$ is taken to be $300$ MPa with 20 $\%$ standard deviation. Our simulations are carried out for material parameters that are reminiscent of aluminum: $E = 70$ GPa, $\nu = 0.33$. We consider $50$ random realizations for each parameter case. If dislocations approach the physical boundary of the sample then a geometric step is created on the surface along the slip direction (see Fig.~\ref{fig:strain_profiles}).

The simulation is carried out incrementally, using a time step that is a factor 20 smaller than the nucleation time $t_{\rm nuc}=10$~ns. At the beginning of every time increment, nucleation, annihilation, pinning at and release from obstacle sites are evaluated. After updating the dislocation structure, the new stress field in the sample is determined, using the finite element method to solve for the image fields~\citep{vandergiessen1995}.

The DIC test that we wish to imitate would measure the strain field in the sample after it has been strained and relaxed, as described above, and then subjected to a subsequent ``DIC-testing'' strain.  (Since we can measure the strain field directly in the simulations, there is no need to simulate the DIC tracking particles.)  We consider a testing reload regime that is governed mainly by the degree of invasiveness we introduce to the data set. All tests have been carried out for prior
loaded states in three different regions (0.1 $\%$, 1 $\%$, 10 $\%$) of total strain. These regions represent samples under different degrees of plastic deformation. Fig.~\ref{fig:schematic2} shows a schematic of the way we create our data set: A sample is loaded to 3 different strains (prior loaded states). From each state we unload to obtain the prior unloaded states. In the prior unloaded states, the samples are stress free, but there is some remaining strain due to plasticity. We then reload the samples to a specific DIC-testing strain. In particular, the strain difference between the test reloaded states and prior unloaded states is constant in all samples. 

\begin{figure}[H] \centering
\includegraphics[width=\textwidth]{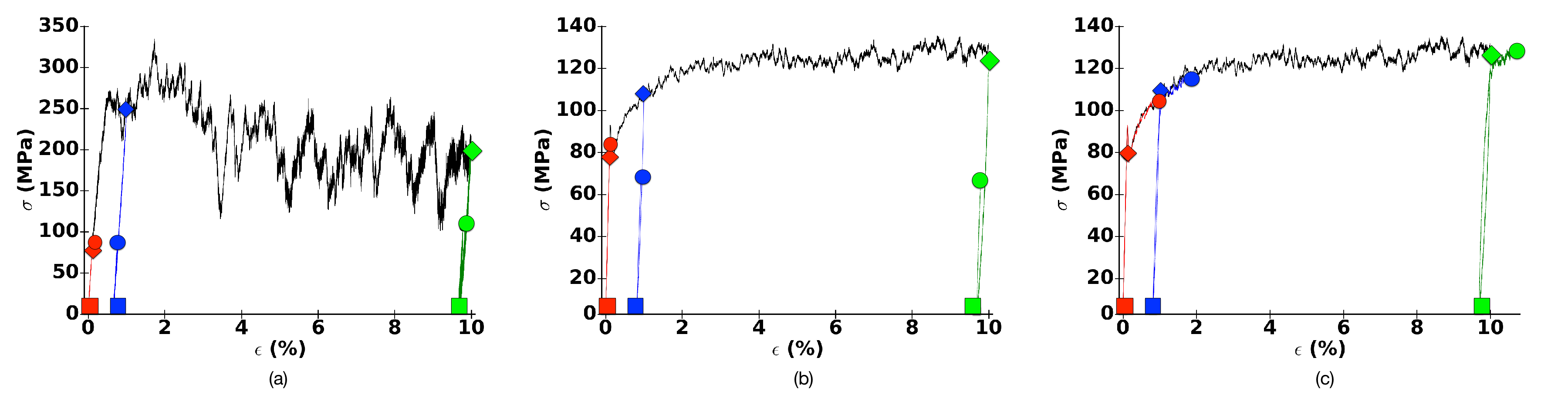}
\caption{\textbf{Examples of stress strain curves obtained from simulations at various widths and DIC-testing strains:} Samples are loaded to 3 different deformations levels (prior loaded states --diamonds). The sample is then unloaded to prior unloaded state (squares). The predeformed sampled is then reloaded to a DIC-testing strain (0.1 $\%$ for $``small-reload"$ and 1 $\%$ for $``large-reload"$ data sets --disks). Each color represents a different loading-unloading-reloading region of the sample. (a) w=0.125 $\micron$, DIC-testing strain = 0.1 $\%$. The prior loaded states are found at 0.1, 1 and 10 $\%$ strain. The prior unloaded states are found at 0.01, 0.59 and 9.61  $\%$ strain. Test reloaded states are at 0.11, 0.69 and 9.71  $\%$ strain. (b) w=2.0 $\micron$, DIC-testing strain = 0.1  $\%$. The prior loaded states are found at 0.1, 1 and 10  $\%$ strain. The prior unloaded states are found at 0.01, 0.76 and 9.73  $\%$ strain. Test reloaded states are at 0.11, 0.86 and 9.83  $\%$ strain. (c) w=2 $\micron$, DIC-testing strain = 1  $\%$. The prior loaded states are found at 0.1, 1 and 10  $\%$ strain. The prior unloaded states are found at 0.01, 0.82 and 9.69  $\%$ strain. Test reloaded states are at 1.01, 1.82 and 10.69  $\%$ strain, respectively.}
\label{fig:stress_curves}
\end{figure}

Samples of different widths ($w$) undergo the same unload-reload protocol to create our data set. We have the option to select at which strain the unload process begins, as well as the DIC-testing strain level we want to introduce.  We perform tests at two values of the reloading strain; the {\em small-reload} data set has a DIC-testing strain of 0.1 $\%$, and the  {\em large-reload} data set has strain of 1.0 $\%$ : that is the difference in strain between prior unloaded states and test reloaded states is 0.1 or 1 $\%$.  Fig.~\ref{fig:stress_curves} shows various characteristic stress-strain curves like that shown schematically in Fig.~\ref{fig:schematic1} (a). The prior loaded state points, as well as the prior unloaded and test reloaded points are shown. Note that there is less noise in the data for larger widths $w$~\citep{papanikolaou2017}. Fig.~\ref{fig:stress_curves} (c) shows a stress strain curve obtained through reloading to larger DIC-testing strain, 1 $\%$. The main difference is at the reload points, which show further deformation of the sample, in contrast to samples reloaded to smaller strain (0.1 $\%$).

\subsection{Emulation of the digital image correlation (DIC) technique for plastically deforming crystals} 
Digital image correlation (DIC) is an optical method that employs tracking and image registration techniques. DIC techniques have been increasing in popularity, including micro- and nano-scale mechanical testing applications, due to its relative ease of implementation and use. Advances in computer technology and digital cameras as well as white-light optics have been instrumental in the development of the technique~\citep{anuta1970,keating1975}. DIC compares a
series of images of a sample at different stages of deformation, tracks pixels movement in a representative volume and calculates displacement and strain with a correlation algorithm. The tracking is done by applying speckled registration dots to the surface of samples. The dots move along as the sample is deformed. Each dot moves both absolutely and relatively to other nearby dots, making possible a reliable identification of the same registration dots in different images, even though the dots have moved. This identification can be used to measure deformation, displacement, strain, and optical flow and it is widely applied in many areas of science and engineering. Through this process we can easily obtain strain fields that may help characterizing a particular elastoplastic
behavior~\citep{anuta1970,keating1975,mccormick2012,roux2009}. DIC only measures relative displacements, it cannot by itself distinguish elastic from plastic strain -- for that, one has to unload the sample and see how much of the strain is recovered.

\begin{figure}[H] \centering
\includegraphics[width=\textwidth]{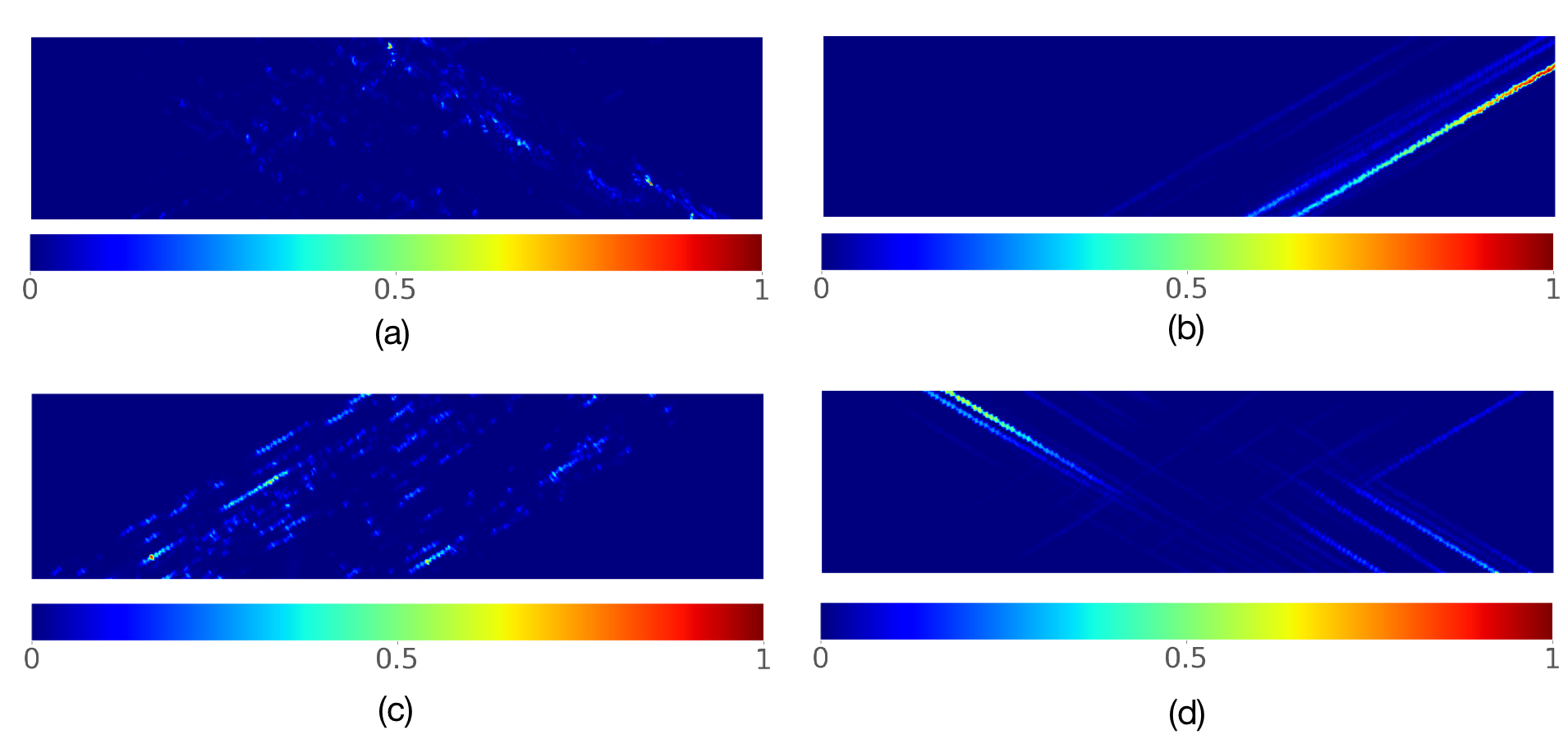}
\caption{\textbf{Variety of strain profiles in 2D-DDD simulations for smaller and larger DIC-testing strain:} A sample is loaded to a high deformation of strain (which could be either 1 $\%$ or 10 $\%$) and then unloaded. This predeformed sample is then reloaded to a DIC-testing strain. In (a),(b) the samples are loaded to 10 $\%$ strain, while in (c),(d) the samples are loaded to 1 $\%$ strain (a) Small DIC-testing strain (0.1 $\%$). w=2 $\micron$. Double slip system simulation. (b) Large DIC-testing strain (1 $\%$). w=1 $\micron$. Single slip system simulation  (c) Small DIC-testing strain (0.1 $\%$). w=1 $\micron$. Single slip system simulation. (d) Large DIC-testing strain (1 $\%$). w=2 $\micron$. Double slip system simulation. For description of color map see Fig.~\ref{fig:strain_profiles} (b).}

\label{fig:deform}
\end{figure}

Our data set consists of samples of various widths under different states of deformation, obtained by simulating thin film uniaxial compression with our 2D-DDD algorithm. Fig.~\ref{fig:deform} shows samples of different widths $w$ under deformation. In particular, (a) and (c) show samples at 0.1  $\%$ total DIC-testing strain while (b) and (d) show samples at 1  $\%$ total DIC-testing strain. These figures are produced through removing the remaining plastic strain (at the prior unloaded states of each sample) from the test reloaded states of each sample. In DIC these figures would be the final images produced after reloading the sample. 

Clearly it is not straightforward to characterize the plastic behavior
of the samples without prior knowledge (i.e. the degree of plasticity
incurred from the prior loaded and test reloaded states). For example,
in Fig.~\ref{fig:deform}, without prior knowledge we would not know
that the samples in (a),(b) are loaded to 10 $\%$ strain while in
(c),(d) the samples are loaded to 1 $\%$ strain. The figures appear to
be quite different, and the similarity of their histories is not
recognizable by eye. Furthermore, (a) and (c) appear to be similar and without prior knowledge, we wouldn't know that the samples were loaded to different strains (the same is true for (b) and (d)). The figures show similarities that only ML's trained eye is able to detect. Indeed, in later sections we will show how ML algorithms can show that the differences in figures are quantifiable and fundamentally different. With the help of ML we are able to find the initial deformation history of various samples, as long as the DIC-testing strain does not overwrite it.

\subsection{Microstructural Discretization}

In this section, we manipulate our microstructural data to make it suitable as input to the subsequent machine-learning steps, closely following the scheme of the Materials Knowledge System (MKS)~\citep{fastkalidindi}

In the general MKS scheme, one selects one or more spatially-varying quantities which characterize the microstructure.  The space of all possible values of these quantities is called the local state space, $\cal{H}$, and a point in this space is denoted $h$. Some care with the vocabulary is required, since physically speaking,
these quantities are simply the values of fields of interest to us, and may or may not correspond to thermodynamic state variables.

In particular, in the current study, we take as our quantity of
interest the determinant of deviatoric total strain (other invariants work equally well) $\phi\equiv \frac{1}{2}(\eps_{xx}^2 + \eps_{yy}^2) -\eps_{xx}\eps_{yy} + 2\eps_{xy}^2$, where the tensor $\eps$ is the total strain, including the plastic part, which is of course not a thermodynamic state variable. The DDD simulations are naturally discretized, so we simply use the DDD grid to index points in space. (See Appendix A-- Fig.~\ref{fig:intro3} for additional details.)

In the MKS method, one further considers a ``microstructure
function'', defined on the product space of the microstructure state
variables $\cal{H}$, and physical space $x$, $m(h,x)$. In general use, this function may be thought of a probability density on these spaces. In our case, where we have a succession of particular microstructural instances, the microstructure function corresponding to each instance is a delta-function in $h$ at each point in space.

In order to obtain data suitable for constructing two-point correlations, it is necessary to bin the state variables.  We make use of the PyMKS software~\citep{wheeler2014} which offers tools to accomplish this.  The most basic $h$-axis discretization scheme is the so-called ``primitive basis'' scheme, in which one selects some number $N$ of evenly-spaced levels, $h_0,h_1,\dots h_N$, and, at a point in space where the state variable has value $h$, selects amplitudes $\omega_i$ for these levels such that $\sum_i{\omega_i h_i}=h$, with the additional restriction that only the $h_i$'s which are directly below and directly above the local value $h$ are nonzero, and $\sum_i{\omega_i}=1$.  The entire system is thus described by a set of values $\{\omega_i\}$ in each spatial bin $x$, from which we can compute auto- and cross-correlation functions $(\omega_i, \omega_i)$ and $(\omega_i, \omega_j)$. In our simulations we discretize the state space into 3 different bins, corresponding to 3 local states $\omega_1$, $\omega_2$, and $\omega_3$ at low, intermediate, and high local strains.

\begin{figure}[tbh] \centering
\includegraphics[width=1\textwidth]{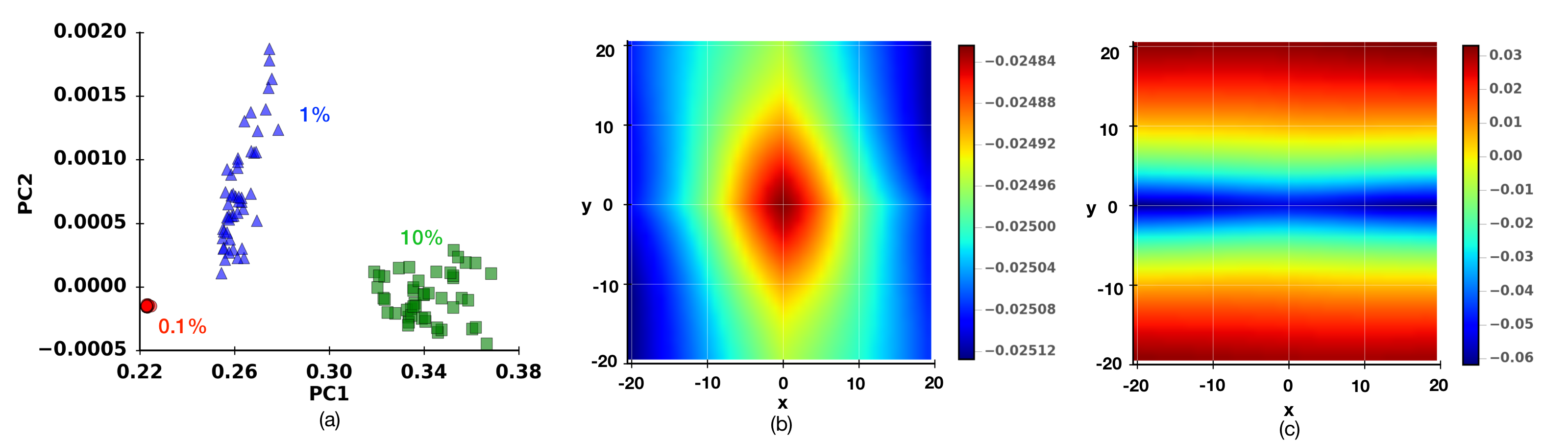}
\caption{ {\textbf{$\boldsymbol{w=2}~\boldsymbol{\mu \rm{m}}$ -- 2D       projection of PCA results for thin films -- Double slip system:}} $\omega_1$,$\omega_1$ auto-correlation.  (a) Projection of data set on first two principal components. Red blobs denote samples with 0.1 $\%$ strain prior loaded state, blue triangles samples with 1 $\%$ strain prior loaded state and green squares denote samples with 10 $\%$ strain prior loaded state, respectively. (b) First principal component of PCA, shown in sample coordinates (Fig.~\ref{fig:schematic}). (c) Second principal component of PCA, shown in sample coordinates (Fig.~\ref{fig:schematic}). The colormaps are unitless.
  % PyMKS's approach to correlation statistics creates correlations
  % limited between values of 0 and 1. Since PCA transforms the data set
  % to uncorrelated variables, then the colormaps are expected to show
  % small deviations in strength around 0. PCA is applied to the
  % correlation matrix and the feature mean is subtracted and this
  % explains the negative values present in some colormaps.
}
\label{fig:width2}
\end{figure}

\begin{figure}[H] \centering
\includegraphics[width=0.82\textwidth]{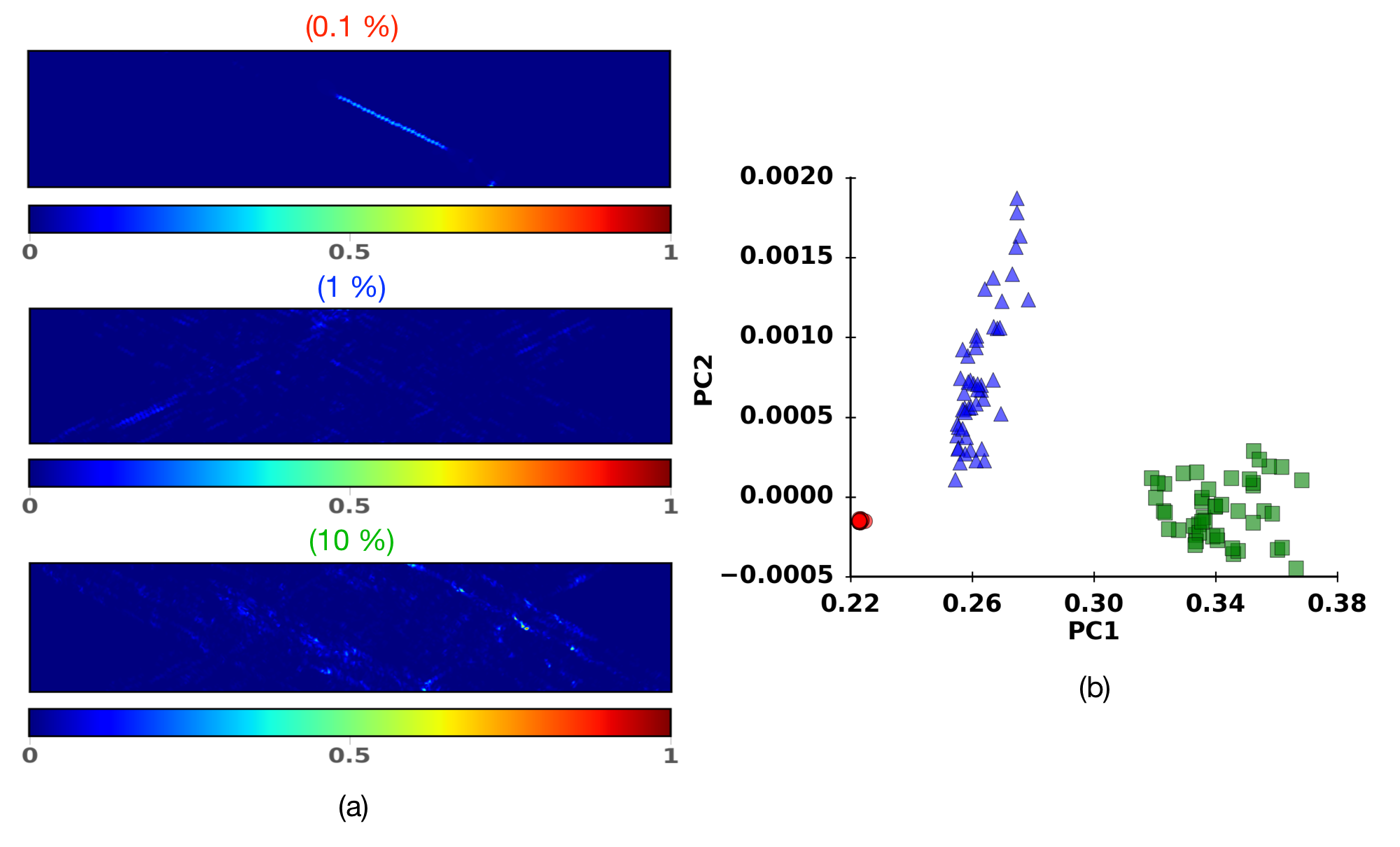}
\caption{ {\textbf{ Efficiency of ML compared to visual inspection of results -- $\boldsymbol{w=2}~\boldsymbol{\mu \rm{m}}$ -- Doubleslip system:}} $\omega_1$,$\omega_1$ auto-correlation. The colors follow the definition of Fig.~\ref{fig:width2}. (a) Strain profiles obtained through our model. The top figure corresponds to strain profiles for a sample initially loaded to 0.1 $\%$ strain. The middle figure is a sample strain profile obtained from initial loading of 1 $\%$ strain. Finally, the bottom figure is a sample strain profile through initial loading at 10 $\%$ strain. All strain profiles have been obtained at the test reloaded state by subtracting the remaining plastic strain at the prior unloaded state. (b) PCA map for samples that have similar strain profiles as (a). 3 distinct clusters are formed , even though samples that were initially loaded to 1 and 10 $\%$ strain have somewhat similar strain profiles (in some areas). The projection is upon the first and second principal components. For description of color map see Fig.~\ref{fig:strain_profiles} (b).}
\label{fig:width2_strain}
\end{figure}

\begin{figure}[H] \centering
\includegraphics[width=0.72\textwidth]{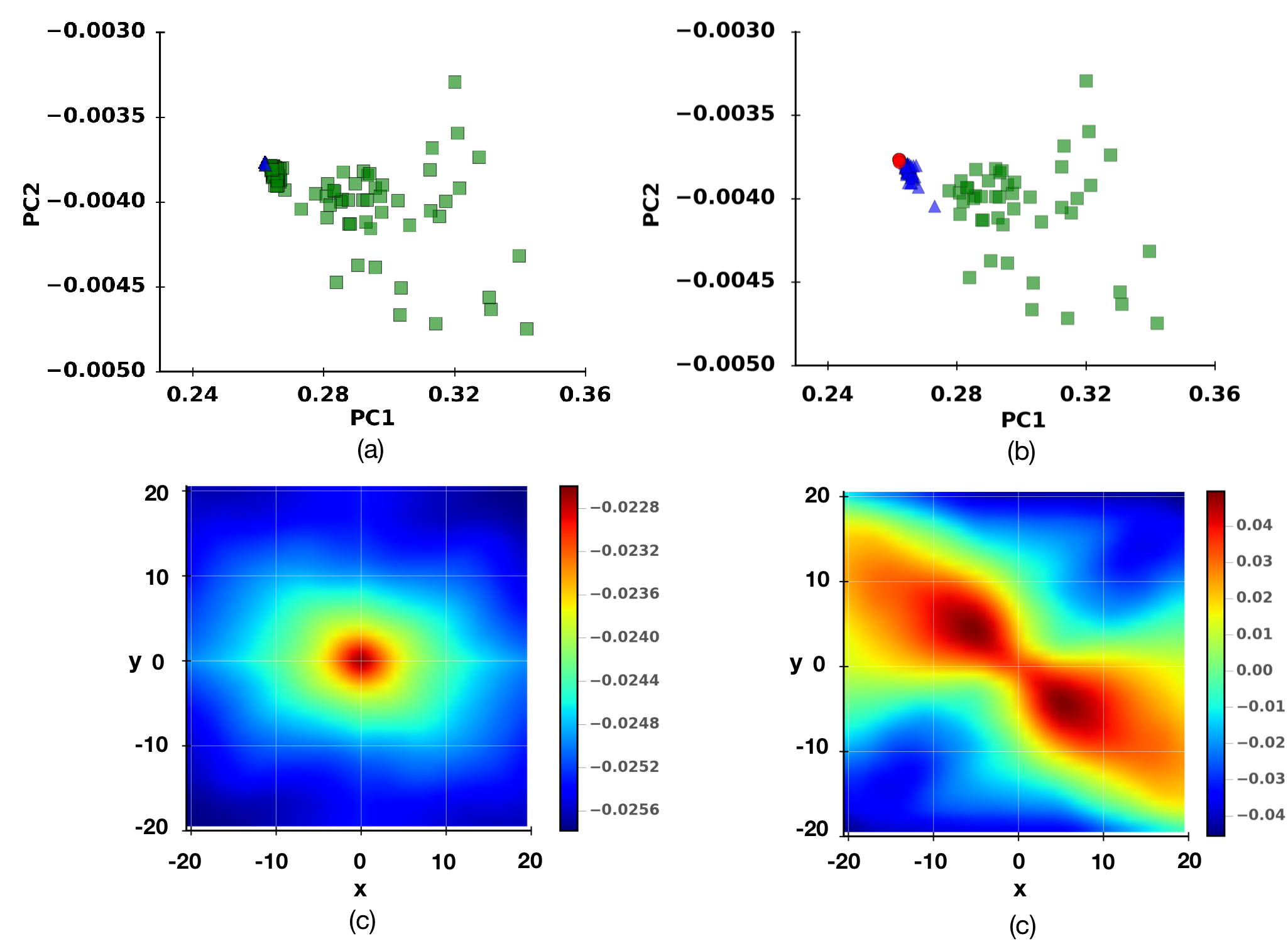}
\caption{ {\textbf{ $\boldsymbol{w=0.5}~\boldsymbol{\mu \rm{m}}$ -- 2D projection of PCA results for thin films -- Double slip system:}} $\omega_1$,$\omega_1$ auto-correlations. The colors follow the definition of Fig.~\ref{fig:width2}. (a) Projection of data set on first two principal components with a clustering algorithm applied to the data set, demonstrating a failure in clustering the various deformation levels. (b) Projection of data set on first two principal components without a clustering algorithm applied to the data set, justifying (a). (c) First principal component of PCA, shown in sample coordinates (Fig.~\ref{fig:schematic}). (d) Second principal component of PCA, shown in sample coordinates (Fig.~\ref{fig:schematic}). For description of colormaps, see Fig.~\ref{fig:width2}.}
\label{fig:width05}
\end{figure}

\section{Pre-processing, clustering and classification for strain profiles of crystalline thin films}

The classification of deformation histories through investigations of strain profiles is not always simple. Certainly, strain localization and shear banding can be strong signatures of local plastic deformation, however with our small DIC-testing strains, localization is not easy to observe. Instead, signatures of plastic deformation are concentrated in collective features of the 2-point correlation function (see Eq.~\ref{2point}). In order to identify and classify these collective features, a statistical approach needs to be implemented in a multitude of training samples. In the following, we use Principal Component Analysis (PCA) as a step in classifying deformation histories.  (There are many other machine learning methods that we could have used instead, but a survey of possible methods lies beyond the scope of this manuscript.) We also discuss the applicability of a classification algorithm to the PCA results.

\subsection{Principal Component Analysis}

After obtaining a set of samples with different plastic deformation histories and calculating statistical correlation functions of each sample, we must show that differences between the correlations can be used to distinguish samples with different ``prior'' plastic strain levels. A handy tool to apply is principal component analysis (PCA). 
 
PCA is a statistical approach that finds an ordered set of orthogonal basis vectors that efficiently describes the variance in a data set \citep{shlens2003}. These vectors are called principal components. The number of principal components is necessarily less than the possible number of observations and the total number of original variables. The basis is defined in such a way that the first principal component has the largest possible variance (that is, accounts for as much of the
variability in the data as possible), and the following components in turn have the highest variance possible, while still being orthogonal to all the preceding components. For this reason, it is critical that all samples in the data set were acquired under identical controlled conditions with a clear understanding of the origin of the variability. PCA is commonly used as one step in a series of analyses; one can use it to reduce the number of variables, especially when there are too many predictors relative to the number of observations.

PCA is accomplished by the application of Singular Value Decomposition (SVD)~\citep{nr}. Given an data matrix, $\matr D$, whose columns each contain the components of a single data point, $D$ may be decomposed to a diagonal matrix of singular values, $\matr S$, and left/right singular vector matrices $\matr V$ and $\matr U$, with $\matr D = \matr V^T \matr S \matr U.$ The columns of $\matr V$ and $\matr U$ are the left and right singular vectors of $\matr D$. The $\matr V$ vectors that correspond to the largest singular values capture the most characteristic features. After calculating the $\matr V$ vectors, we project the data samples onto the subspace defined by them.  In decreasing order of their corresponding singular values, we denote the first three basis vectors (principle components) as PC1, PC2, and PC3.

In Machine Learning, the PCA technique is used to project data  vectors which lie in a high-dimensional space to a smaller dimensional space whose basis vectors are selected to capture most of the variation in the original data. In the textbook scheme, one typically has an $n\times m$ array, with $n$ indexing the samples, and $m$ being the dimensionality of the feature space the vectors lie in. In our case, we wish to do PCA on our correlation data, which we can treat as a vector by listing the data values from the grid in raster order. 

For any given sample, it is possible to compute the two-point correlations out to any distance that can be supported by the data, but in order to meaningfully compare correlations across samples of different sizes, it is necessary to have uniform input data to our eventual machine-learning pipeline. For this reason, we truncate the two-point correlation vector on to a square grid that we believe is large enough to be a useful statistical characterization of the spatial arrangement of the strain field, but small enough to give reasonable execution speeds in our numerical analysis operations.

Having projected the correlation data to the PCA-derived subspace, we can address our main objective, to find the prior deformation history of the samples from the strain correlation measurements. We generate data on samples of different widths w (0.125, 0.25, 0.5, 1.0, 2.0 μm) for both single and double slip systems, reloaded to two values of the DIC-testing strain which we call ``large-reload" (1 $\%$) and ``small-reload" (0.1 $\%$). The aspect ratio is fixed ($h/w$ $=$ $4$), and the number of pixels in each of the DDD simulations is fixed to 80 (in both dimensions). We calculate the two-point correlation matrix for all samples of each width and restrict the data to the 40 by 40 region. Continuing, we adjust the truncated correlation matrix by subtracting the average.  The end result is used as an input for the start of our ML method. The information is fed to the PCA algorithm in order to change from the high dimensional space of the correlation matrix, to a more controlled subspace of the principal components. In our calculations the first three components of PCA describe more than 98 $\%$ of the variance in the data sets, so we restrict the computation to those three components. Next, we project the correlation matrix onto the new reduced subspace. Visually, at this point, we are able to examine the possible clusters that have formed due to the different PCA vectors that have been produced. Furthermore, we are able to identify the samples that belong to each cluster, which we use as an input for examining the accuracy of classification algorithm. The set of points belonging to the 3D subspace of PCA (PC1,PC2,PC3) are used as an input to the classification algorithm to examine whether or not the visually identified clusters are replicated by the classification algorithm.

\begin{figure}[H] \centering
\includegraphics[width=1\textwidth]{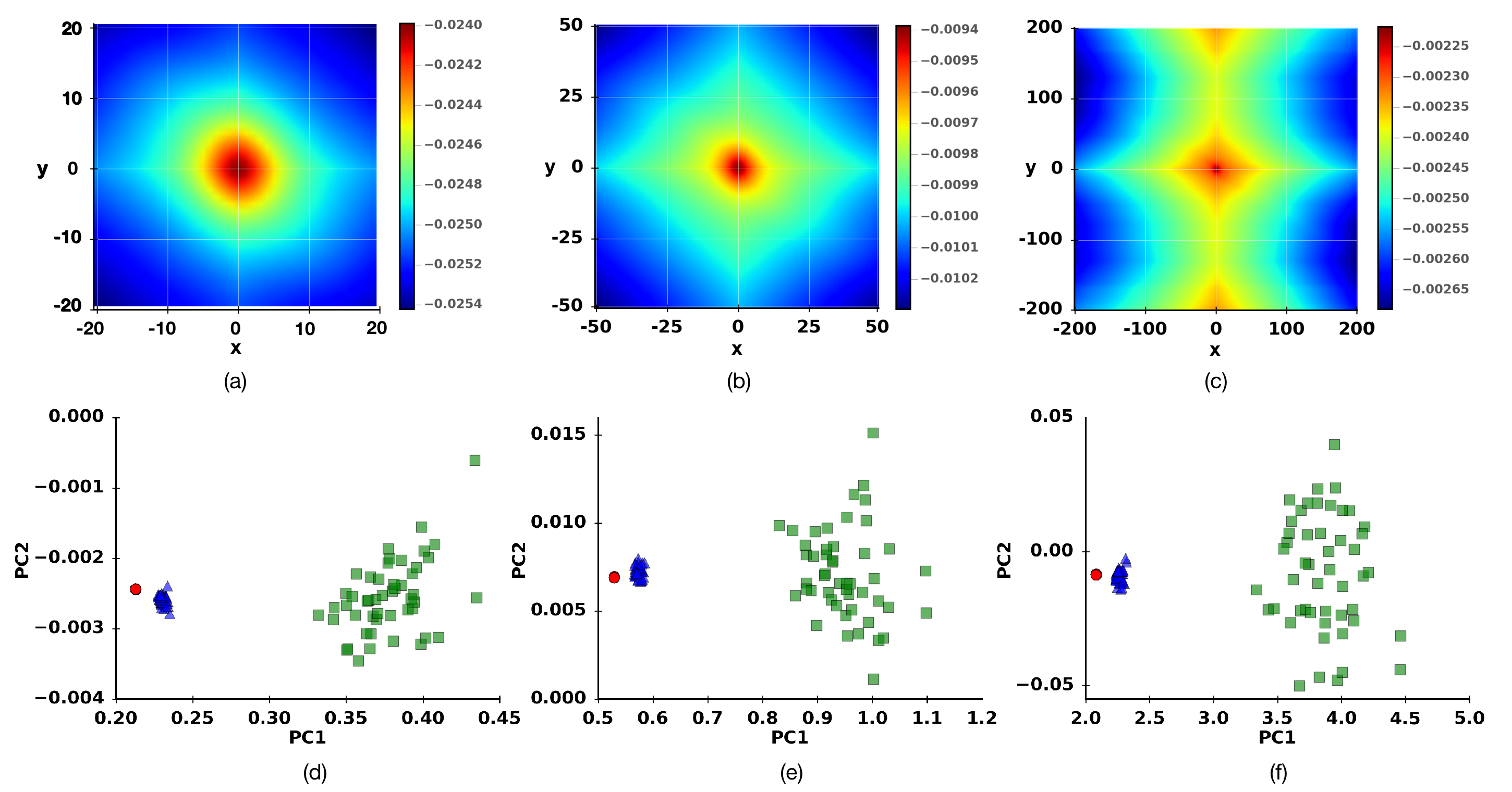}
\caption{ {\textbf{$\boldsymbol{w=1}~\boldsymbol{\mu \rm{m}}$, The choice of the correlation domain and how it impacts the PCA maps -- Double slip system :}} $\omega_1$,$\omega_1$ auto-correlation. For description of colormaps and colors of PCA maps, see Fig.~\ref{fig:width2}. Projection of data set on first two principal components. (a) 40 by 40 domain of correlation matrix. Highly smooth in the center and towards the boundaries of the domain. (b) 100 by 100 domain of the correlation matrix. A highly focused area near the center of the domain is shown, where the phenomena are focused. The smoothness present in (a) is slowly removed from this domain. (c) 400 by 400 domain of correlation matrix. Along the y-axis we have hit the actual boundaries of the sample. We have rich phenomenology present towards the center of the correlation matrix and at the boundaries. (d) PCA maps for 40 by 40 domain. (e) PCA map for 100 by 100 domain. The variance of the data has changed and the projections have shifted. The information provided by (b) doesn't change the cluster formations, but introduces unnecessary information that has shifted the results along the PC1 and PC2 axes. (f). PCA map for 400 by 400 domain. The variance of the data has changed even more compared to (e). The distances between the blue and green clusters have increased an order of magnitude compared to (e) and 2 orders of magnitude compared to (d). The information provided by (c) doesn't affect the clusters that are formed from our algorithm.}
\label{fig:progression}
\end{figure}

\subsection{Clustering and classification}

We use the Continuous k-Nearest Neighbors (CkNN) algorithm~\citep{berry2017} to classify samples after running PCA on the data set. CkNN searches for samples that are close to each other (k-nearest neighbors) and arranges them into clusters.  The algorithm is an unsupervised method that detects natural clusters within a data set, and our interest in it is the degree to which the natural clusters correspond to the prior deformation. After different clusters have been identified, the algorithm classifies the samples based on the cluster to which they belong. The input to this algorithm must consist of a set of points, which in our case is the projections of the correlation matrix on the 3 principal components. As an output, the algorithm produces the classified samples, based on the cluster to which they belong. 

The size of the data set, as in most classification algorithms, imposes a limitation on the algorithm. For double slip system simulations our data set consists of 150 samples for each $w$ while for single slip system simulations our data set consists of 27 samples for each $w$. This introduces a limitation on classification, especially for single slip systems. The algorithm groups data samples with similar PCA vectors into one cluster. We find that the algorithm works better for larger data sets. The method is successful if the samples with different prior loading are grouped into different clusters. Note that the clustering is done in three dimensions using all three PC components, but most of our plots are two dimensional, which can sometimes hide the degree of clustering. 

\begin{figure}[H] \centering
\includegraphics[width=0.62\textwidth]{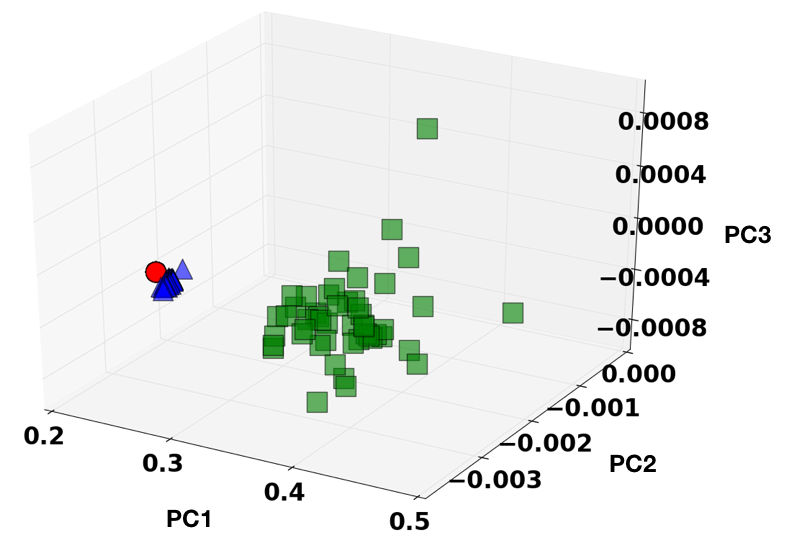}
\caption{ {\textbf{$\boldsymbol{w=1}~\boldsymbol{\mu \rm{m}}$, 3D projection of PCA results for thin films -- Double slip system :}} $\omega_1$,$\omega_1$ auto-correlation. The colors follow the definition of Fig.~\ref{fig:width2}. 3 different clusters are shown like in Fig.~\ref{fig:width1}. Introducing the 3rd component into the PCA map, doesn't affect the results.}
\label{fig:3dplot}
\end{figure}

In Fig.~\ref{fig:3dplot} we show a 3D PCA map for material samples of $w = 1 \micron$. It is obvious that the clustering isn't affected by Mthe 3rd dimension, and in this case the information provided by PC3 is irrelevant to our results. 

The results shown on this paper, except for results shown in section 3.5, are extracted by applying the PCA and the CkNN algorithm to the whole data set. The same PCA and CkNN steps are applied to all the simulations. The remainder of this paper discusses how well the clustering algorithm works in various situations.

\begin{figure}[H] \centering
\includegraphics[width=0.72\textwidth]{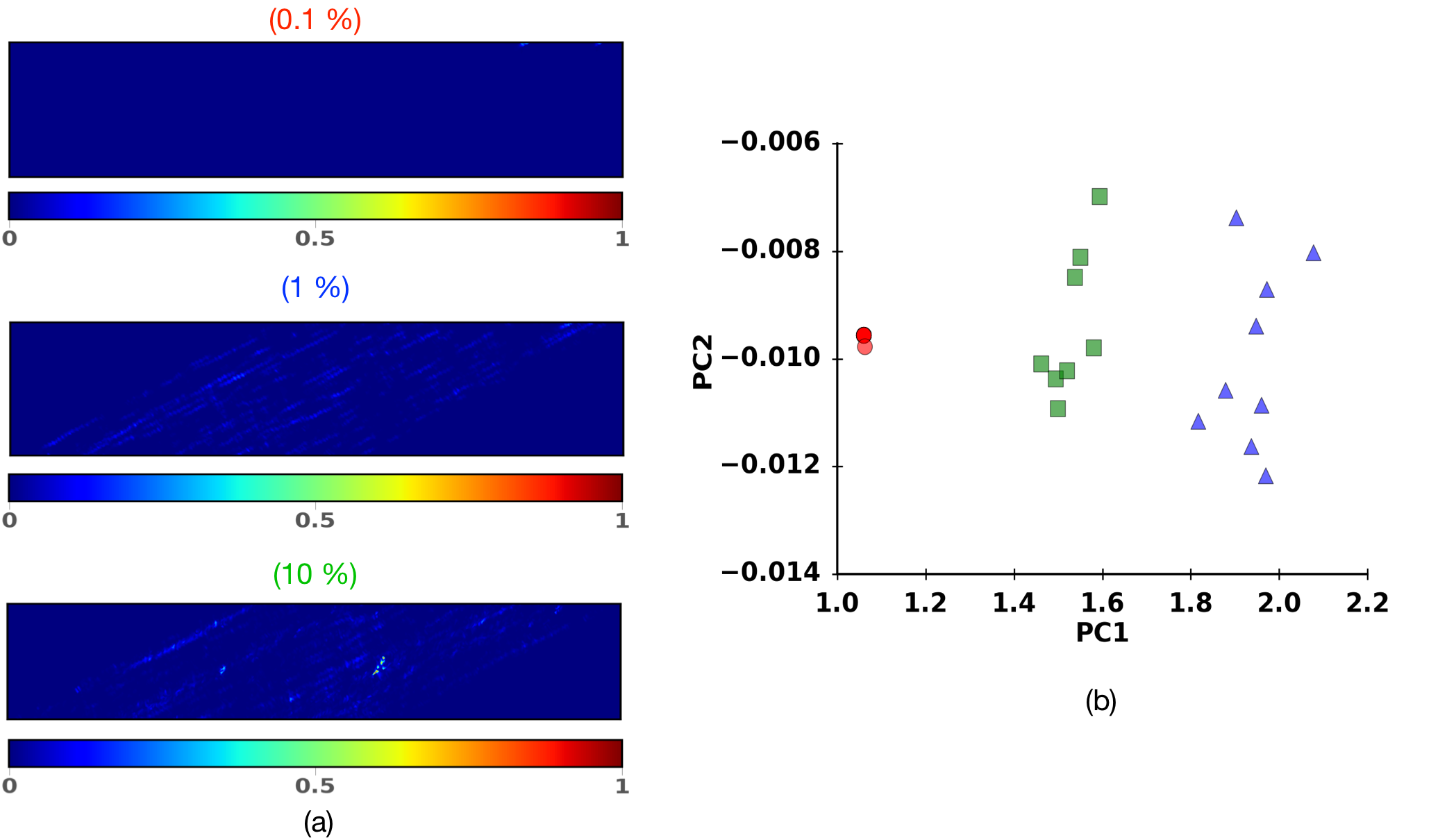}
\caption{ {\textbf{$\boldsymbol{w=2}~\boldsymbol{\mu \rm{m}}$,
-- Strain profiles and 2D projection of PCA results for thin films -- Single slip system :}} $\omega_1$,$\omega_1$ auto-correlation. The colors follow the definition of Fig.~\ref{fig:width2}. Projection of data set on first two principal components. (a) Different strain profiles are seen for a single slip system of $w=2 \micron$. These strain profiles are obtained through subtracting the remaining plastic strain at the prior unloaded state, from the test reloaded state. The top profile is for sample initially loaded to 0.1 $\%$ strain, then unloaded to zero stress and reloaded to DIC-testing strain of 0.1 $\%$. The middle profile is for sample initially loaded to 1 $\%$ strain and then unloaded and reloaded. The bottom figure is for sample initially loaded to 10 $\%$ strain and then underwent the unload-reload process. (b) PCA projection of of our results for samples of $w=2$ $\micron$. The similarity between strain profiles at 1 and 10 $\%$ doesn't affect the formation of separate clusters for samples that were initially loaded at these strains. For description of color map see Fig.~\ref{fig:strain_profiles} (b).}
\label{fig:width2-1-oneslip}
\end{figure}

\begin{figure}[H] \centering
\includegraphics[width=0.72\textwidth]{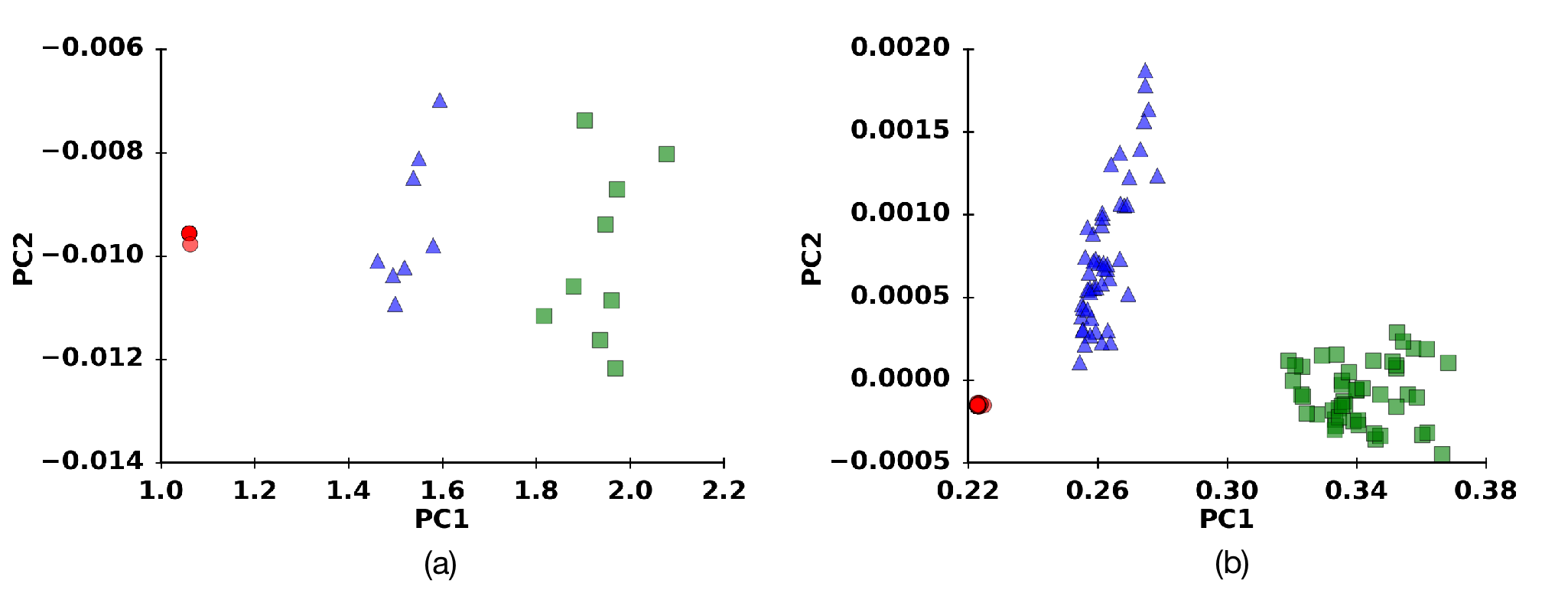}
\caption{ {\textbf{$\boldsymbol{w=2}~\boldsymbol{\mu \rm{m}}$ -- 2D projection of PCA results for thin films -- Comparison between single and double- slip system :}} $\omega_1$,$\omega_1$ auto-correlation. The colors follow the definition of Fig.~\ref{fig:width2}. Projection of data set on first two principal components. (a) Single slip system projection (b) Double slip system projection.}
\label{fig:width2_comparison_double_single}
\end{figure}

\begin{figure}[H] \centering
\includegraphics[width=0.72\textwidth]{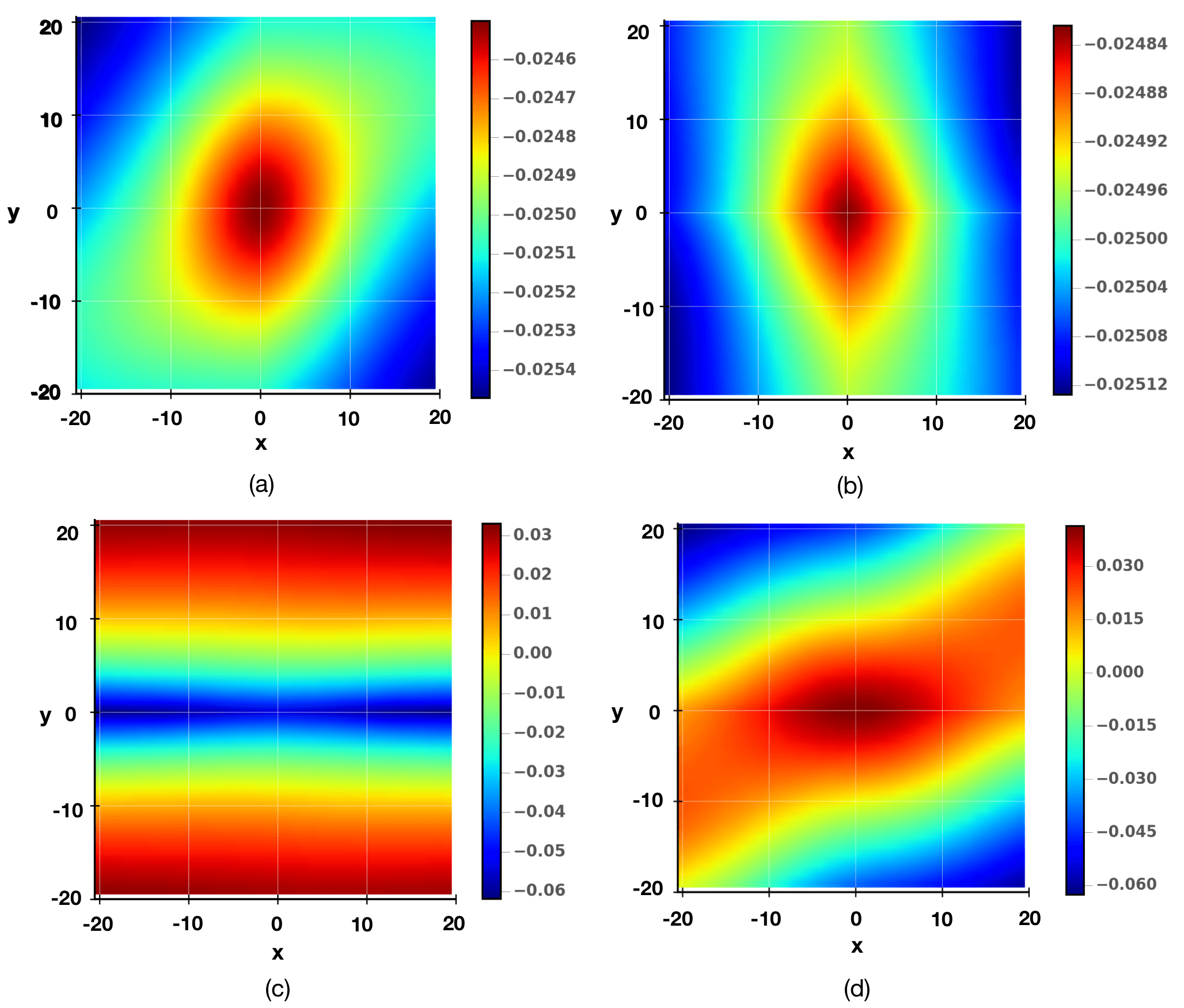}
\caption{{\textbf{$\boldsymbol{w=2}~\boldsymbol{\mu \rm{m}}$ -- Comparison of principal components among double and single slip systems:}} Components shown correspond to the analysis of Fig.~\ref{fig:width2_comparison_double_single}. (a) First principal component of double slip system. (b) First principal component of single slip system.  (c) Second principal component of double slip system. (d) Second principal component of single slip system. For description of colormaps, see Fig.~\ref{fig:width2}.}
\label{fig:components_double_single}
\end{figure}

\newpage
\subsection{Distinguishing plasticity regimes for small DIC-testing strain (0.1 $\%$)}

We ran a multitude of tests for different $w$.  For large $w$ ($>0.5 \micron$), our algorithm correctly clusters and classifies data into 3 different groups, one for each of the prior strain values, which was
the main objective of our work.
%In the following figures, we consider
%auto-correlations between local states $\omega_1$ and $\omega_1$.
Fig.~\ref{fig:width2} (a) shows that clustering is easily observed for $w=2\micron$, where 3 distinct clusters appear in the PCA of the m$\omega_1$ autocorrelation.  It is clear that there is enough cluster separation to reliably classify plastically deformed metals into heavily deformed and less deformed categories. For these larger sized systems the CkNN algorithm has 100 $\%$ accuracy, while for smaller sized systems with $w \leq 0.5 \micron$ the clustering algorithm fails to cluster data points according to their deformation state. That is evident in Fig.~\ref{fig:width05} (b), where one can see what a correct clustering and classification would look like for specimens of $w=0.5\micron$. In Fig.~\ref{fig:width05} (a) one can observe the results after the CkNN algorithm is applied to the data set.  Other figures in the Appendix C.1 show how specimens of various sizes are classified using the CkNN algorithm. The plastic noise fluctuations in the system, as well as the finite size of the system, interferes with the classification of smaller sized data samples, while for larger $w$ the samples are classified correctly. 

Fig.~\ref{fig:width2} (b) and (c) show the representation of the first 2 principal components, shown in their natural sample coordinates ({\it i.e.}, the PCA vectors have been converted back to the 2D grid representation of a correlation function) with units denoting number of elements (in regular DIC, it would correspond to number of pixels in each direction). If two correlation functions were randomly chosen from the data set, the difference between them would most likely look like \ref{fig:width2} (b) (with some scaling) mixed with a smaller amount of \ref{fig:width2} (c). Note that the first principal component is roughly isotropic, while the second is strongly anisotropic. Figs.~\ref{fig:width2}, \ref{fig:width05} show the progression of PCA's effectiveness as a clustering technique as sample width decreases. We can observe that the first PCA component at larger $w$ is relatively isotropic. While in Fig.~\ref{fig:width2}~(b) we notice a concrete isotropy of the first principal component of the analysis, it gradually becomes anisotropic as the sample width decreases (Fig.~\ref{fig:width05} (b)). This change is correlated with the onset of stochastic fluctuations at small scales and mechanical annealing~\citep{shan2008} that promotes concrete slip bands even at small DIC-testing strains. While both principal components for $w=2 \micron$ (Fig.~\ref{fig:width2} (b,c)) are smooth, they gradually become less structured as $w$ decreases (Fig.~\ref{fig:width05} (b,c)), naturally an effect of stochastic fluctuations at small length scales. For $w=2$ $\micron$ there is a distinct difference between the first and second principal components, related to a spatial symmetry breaking. This distinction disappears as $w$ decreases. For smaller $w$, due to the emerging crystal plasticity size effects~\citep{papanikolaou2017}, the data set is not as distinguishable as we would have wanted with our clustering technique, because of the noise associated with strengthening. (Fig.~\ref{fig:stress_curves} (a)). Examples may be seen in Fig.~\ref{fig:width05} and in Appendix C.1. These figures show the behavior of our classifier and how they compare to the actual initial deformation of our samples.

One deficiency of our procedure emerged as we examined the results: as $w$ decreases, the distance between the PCA-transformed samples also decreases. It is known that classification algorithms have an inherent limitation: when the distance between points in one cluster is similar to the distance separating two clusters, then the algorithm has difficulty distinguishing the clusters. In particular, Fig.~\ref{fig:width2} shows that the cluster distances in the PC1 direction are of order of magnitude $10^{-2}$ to $10^{-1}$. For $w \leq 0.5\micron$ the cluster between PCA-transformed samples is on the order of $10^{-3}$ to $10^{-2}$, similar to the distance between the samples itself, and the data samples cannot be classified correctly. For smaller sized systems, it is evident that samples with 0.1 $\%$ and 1 $\%$ prior loaded strain (red circles and blue triangles, respectively) are so close to each other that the classifier regards them as belonging to the same cluster.

\subsection{Distinguishing plasticity regimes for single slip samples}

As mentioned in Sec.~2.1, we model single and double slip systems. So far, we have shown how emergent shear bands can be observed in our simulations for both of these systems (Fig.~\ref{fig:deform}), as well as PCA results for double slip (Figs~\ref{fig:width2},~\ref{fig:width05}).  PCA results for single slip are consistent with double slip, as shown in Fig.~\ref{fig:width2-1-oneslip}.  Specifically, Fig.~\ref{fig:width2-1-oneslip} (a) shows results of single slip system simulations for $w=2$ $\micron$ and (b) for $w=1$ $\micron$. The clustering properties for these larger sized systems are similar to the properties observed for similar systems for double- slip simulations. Fig.~\ref{fig:width2_comparison_double_single} displays a direct comparison between the results of single slip and double slip systems with $w=2$ $\micron$.  The PCA results contain distinctly separated clusters. 

Fig.~\ref{fig:components_double_single} compares the principal components of the structure factors for single and double slip systems.

\begin{figure}[H] \centering
\includegraphics[width=0.52\textwidth]{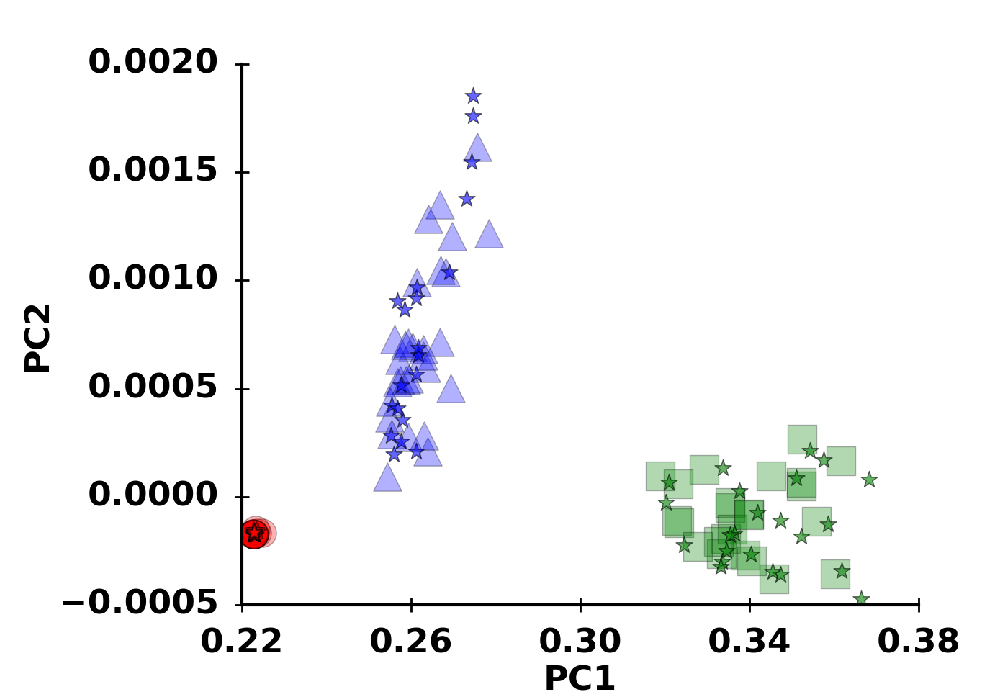}
\caption{ {\textbf{$\boldsymbol{w=2}~\boldsymbol{\mu \rm{m}}$ -- 2D projection of PCA results for thin films -- Double slip system -- Validation:}} $\omega_1$,$\omega_1$ auto-correlation. Red blobs denote $training$ samples with 0.1 $\%$ strain prior loaded state, blue triangles $training$ samples with 1 $\%$ strain prior loaded state and green squares denote $training$ samples with 10 $\%$ strain prior loaded state, respectively. Red stars depict $testing$ samples of 0.1 $\%$ prior loaded state, blue stars $testing$ samples of 1 $\%$ prior loaded state and green stars $testing$ samples of 10 $\%$ prior loaded state. Validated-split data set. Projection on first two principal components.
}
\label{fig:validation}
\end{figure}

\subsection{Validation and accuracy of the algorithm}

A ML algorithm, in order to be considered successful, should be validated with ``unknown" data sets (testing data) which have the same features as the data set the algorithm was designed for (training data). In many cases, testing data sets are hard to find, so the whole data set is split into two parts (not necessarily a half and half split), and the ML algorithm can be trained on part of the data set and its effectiveness tested on the rest.
The results shown in the rest of this paper are an application of PCA transformation and the CkNN algorithm on the whole data set (for a given $w$). 

For validation purposes, we ``trained'' the algorithm by computing the PCA transformation from a randomly chosen half of the $w=2$ $\micron$ samples and applying the CkNN algorithm. Then we applied the PCA transformation to the remaining half of the samples and examined whether or not they were projected into the correct clusters. The results are shown in Fig.~\ref{fig:validation}. It is evident that the testing data perfectly matches the training set. Similarly ``training" the algorithm to samples of various sizes (i.e., half of the samples instead of all the samples), follows the results of section 3.3. For samples with $w$ $\geq$ $1$ $\micron$ the ``testing" data set is projected to the 3 classified clusters that have formed. In contrast, for smaller sized systems, the training data set is misclassified (as happens when examining the whole data set) and the testing data set falls within the misclassified results. 

\begin{figure}[H] \centering
\includegraphics[width=1.0\textwidth]{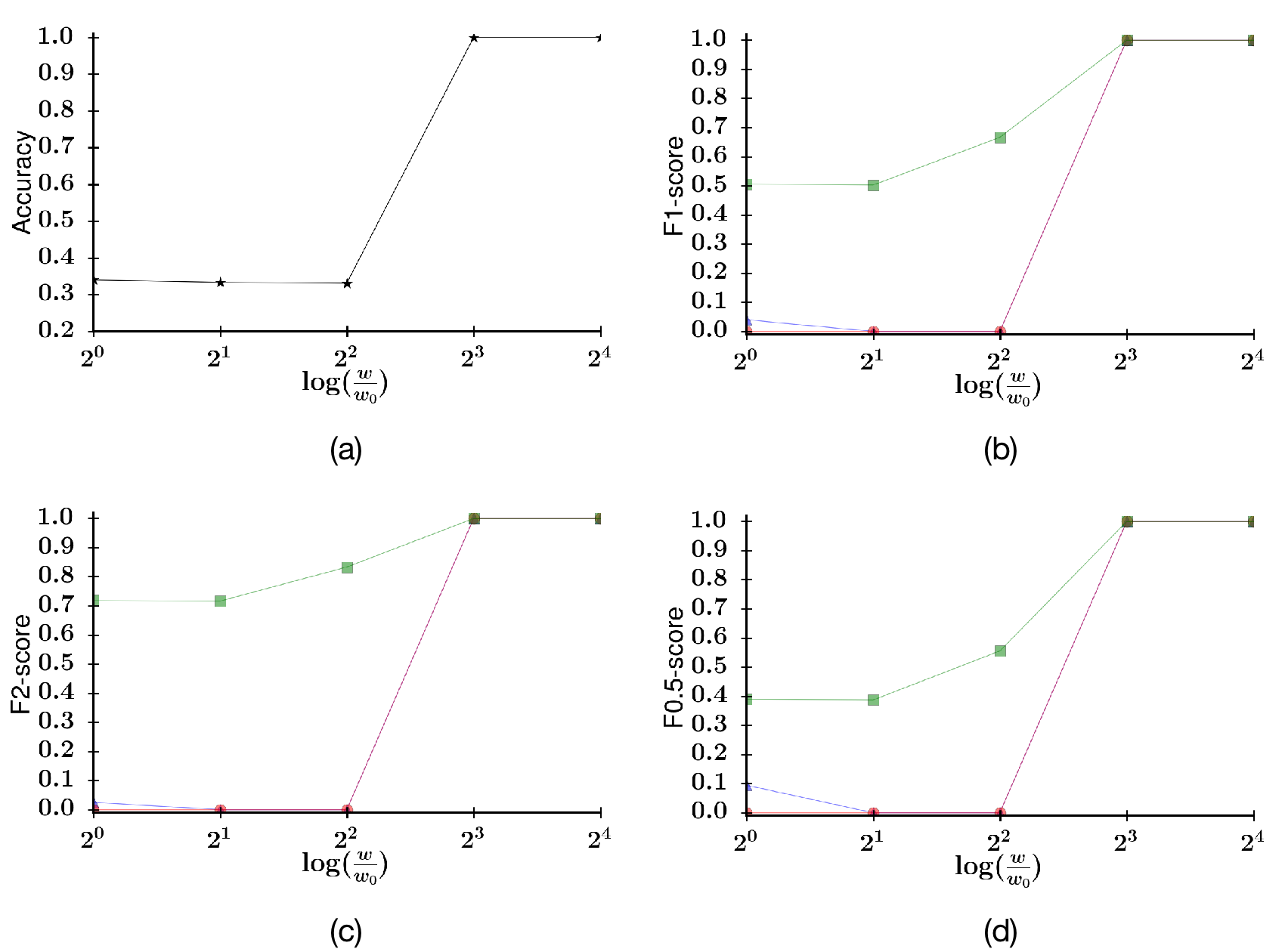}
\caption{ {\textbf{Accuracy scores of classified samples - 0.1 $\%$ DIC-testing strain:}}. $\omega_1$,$\omega_1$ auto-correlations. The x-axis of each graph is the base 2 logarithm of the various sample widths we examine. (a) Accuracy score for the samples. Maximum value 1 means that all the samples have been correctly classified. (b) $F_1$-score of our 3 clusters that are formed. The line with the squares represents the cluster with samples of 10 $\%$ initial strain loading, while the line with the triangles is for the cluster with samples initially loaded to 1 $\%$ strain. Finally, the line with the circles is for the cluster with samples initially loaded to 0.1 $\%$ strain. For smaller sized systems we have observed that most of the samples are classified as belonging in the ``square" cluster, hence the scored value for that cluster only. Since the algorithm correctly classifies the samples that were initially loaded to 10 $\%$ strain, but also classifies more samples as belonging to that cluster, then the score does not have the maximum value of 1 but lower. (c) $F_2$-score of our 3 cluster that have formed. The definition of the colored lines follows (b). Since for F2-score we have increased weight of the recall, the 0.7 maximum value is expected for the ``square" cluster. (d) $F_{0.5}$-score of our 3 clusters. The colors definitions follow (b). Since we have reduced weight of the precision, for lower sample widths it is expected to have lower score than $F_1$ for the ``square" cluster.}
\label{fig:accuracy}
\end{figure}

% ACR Item 8. 
We can quantify the degradation of the clustering process using some of the tools provided in the scikit-learn metrics module~\citep{pedregosa2011}. In particular, we examine the accuracy score of the algorithm, as well as the $F_{\beta}$ score. Accuracy is the ratio of the number of classified samples by our algorithm (prediction) over the number of the true classification of our samples. 
{ We apply the CkNN algorithm and generate clusters. Because we know the prior strain  for each sample, we can immediately check whether the clusters correspond to the strain levels. Perfect clustering is when each cluster contains only samples with identical prior strains.} The results are summarized in Fig.~\ref{fig:accuracy}. For $w$ $\geq$ $1$ $\micron$ the accuracy score is 1 as seen in Fig.~\ref{fig:accuracy} (a); that is, all the samples are correctly classified. For smaller samples $w$ $\leq$ $0.5$ $\micron$ (or ${w}/{w_0}$ $\leq$ $2^2$ as in the figure), we have a 0.33 accuracy score, because only the samples of one cluster are correctly classified. The CkNN algorithm predicts that most samples belong to that cluster, but this doesn't affect the accuracy score, since the samples that belong to that cluster are correctly classified.

To quantify the performance of the classification process, we use the
$F_{\beta}$ score~\citep{powers2011,yates2011}, 

% \begin{equation}
% F_{\beta} = (1+\beta^2)\cdot\frac{precision\cdot recall}{(\beta^2\cdot precision) + recall}
% \end{equation}

\begin{equation}
 F_{\beta} = (1+\beta^2)\cdot\frac{p\cdot r}{(\beta^2\cdot p) + r}
\end{equation}

where precision $p$ is the number of correctly classified samples in the cluster divided by the number of all classified samples in the same cluster, and recall $r$ is the number of correctly classified samples in the cluster divided by the number of samples that should have been returned (classified). The $\beta$ number changes the weight of recall $vs$ precision. For $\beta$ $>$ 1 then recall is weighted more than precision, while for $\beta$ $<$ 1 precision is weighted more than recall. For $\beta$ $=$ 1 then we have the $F_1$-score, with precision and recall having the same weight in the equation. In our work, we examine the $F_1$, $F_2$ and $F_{0.5}$-score. In Fig.~\ref{fig:accuracy} (b,c,d) we see the respective results for these scores. For samples with $w$ $\geq$ 1 $\micron$ (or ${w}/{w_0}\geq 2^3$ we have value of 1 on all scores and all clusters, while for smaller $w$ we observe that the line with the squares, which corresponds to samples with 10 $\%$ initial compressive loading returns non-zero values, varying as the $\beta$ value changes. For the samples belonging to that cluster we don't get the highest possible result, because the number of correctly classified samples is smaller than the number of samples in the cluster (i.e., the precision is small). The line with the circles, which corresponds to samples with 0.1 $\%$ initial strain loading, has value 0 for $w$ $\leq$ 0.5 $\micron$ because no samples have been classified as belonging to that cluster. The last line, with the triangles corresponding to samples with 1 $\%$ initial loading has non-trivial values because in some cases there are some samples that are classified correctly (the recall and precision are very small). In summation: For the ``square" cluster we have low precision but high recall,since we classify the samples that actually belong that the cluster correctly, but we also classify samples from other clusters; for the ``triangle cluster" we have low recall and low precision, since we classify a small number of samples into that cluster.

\begin{figure}[H] \centering
\includegraphics[width=0.72\textwidth]{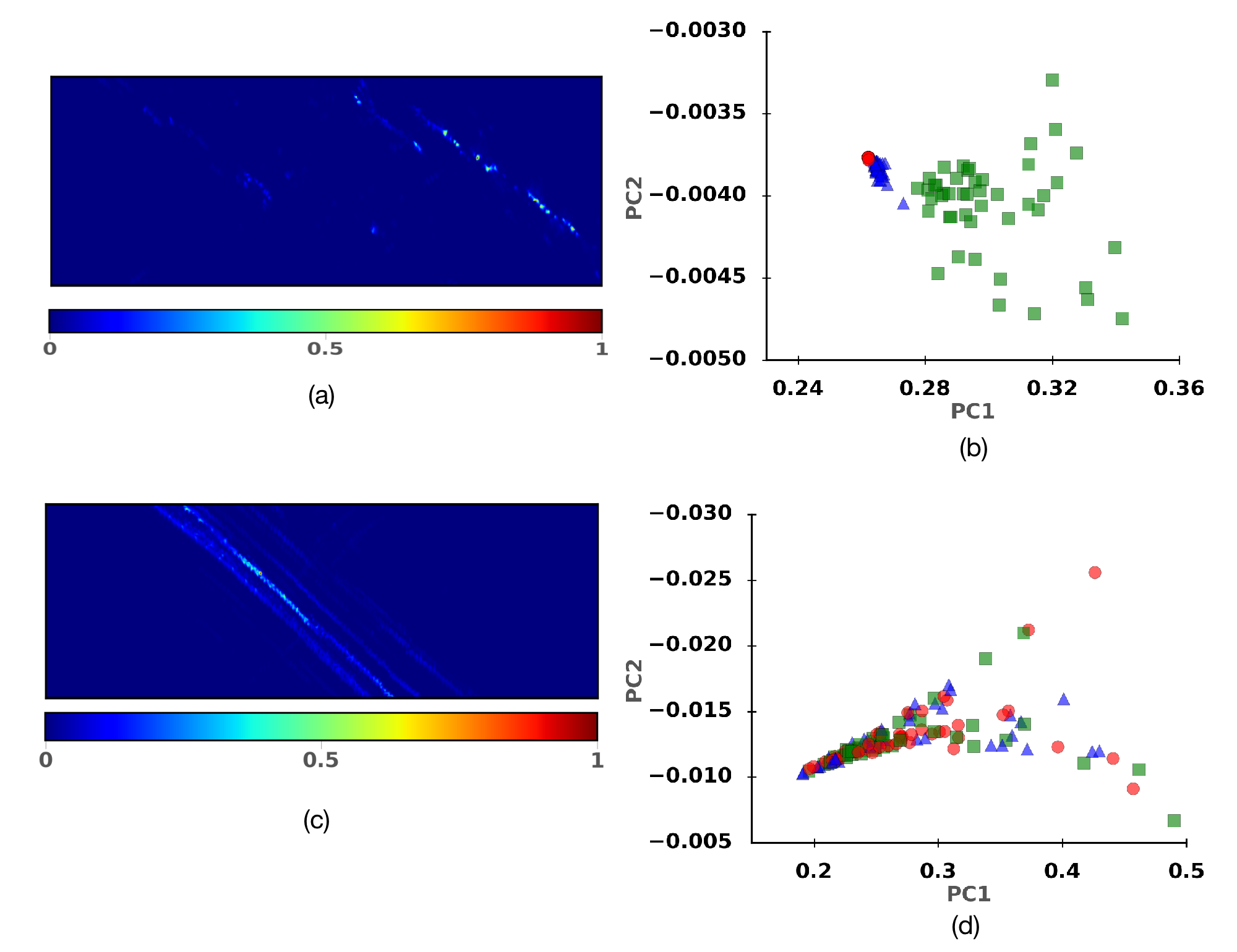}
\caption{ {\textbf{Large-reload {\it vs.} small-reload DIC-testing -- Example of PCA projection results for thin films of $\boldsymbol{w=0.5}~\boldsymbol{\mu \rm{m}}$:}}. $\omega_1$,$\omega_1$ auto-correlations. The colors follow the definition of Fig.~\ref{fig:width2}. (a) Sample with prior loaded state at 10 $\%$ strain is unloaded to zero stress and then reloaded to small DIC-testing strain (0.1 $\%$). The strain profile shown is captured after removing the prior unloaded state strain profile. (b) Small DIC-test reloaded state strain (0.1 $\%$), without a clustering algorithm applied to data set. Projection on two principal components. Actual representation of the data set, with some mixing of the samples. The clusters have shifted closer to one another but not indistinguishable. (c) Sample with prior loaded state at 10 $\%$ strain is unloaded to zero stress and then reloaded to large DIC-testing strain (1 $\%$). The strain profile shown is captured after removing the prior unloaded state strain profile. The difference in strain profiles (a) and (c) is clearly shown. (d) Large DIC-test reloaded state (1 $\%$) without a clustering algorithm applied to the data set. Actual representation of the data set. For the higher DIC-testing strain of 1 $\%$, we can see that there is much more mixing of the samples. Reloading to higher strain values, adds plastic memory to the samples, rendering our process inapplicable for these cases. For description of color map see Fig.~\ref{fig:strain_profiles} (b).}
\label{fig:inv-noninvcomp}
\end{figure}

\begin{figure}[H] \centering
\includegraphics[width=0.72\textwidth]{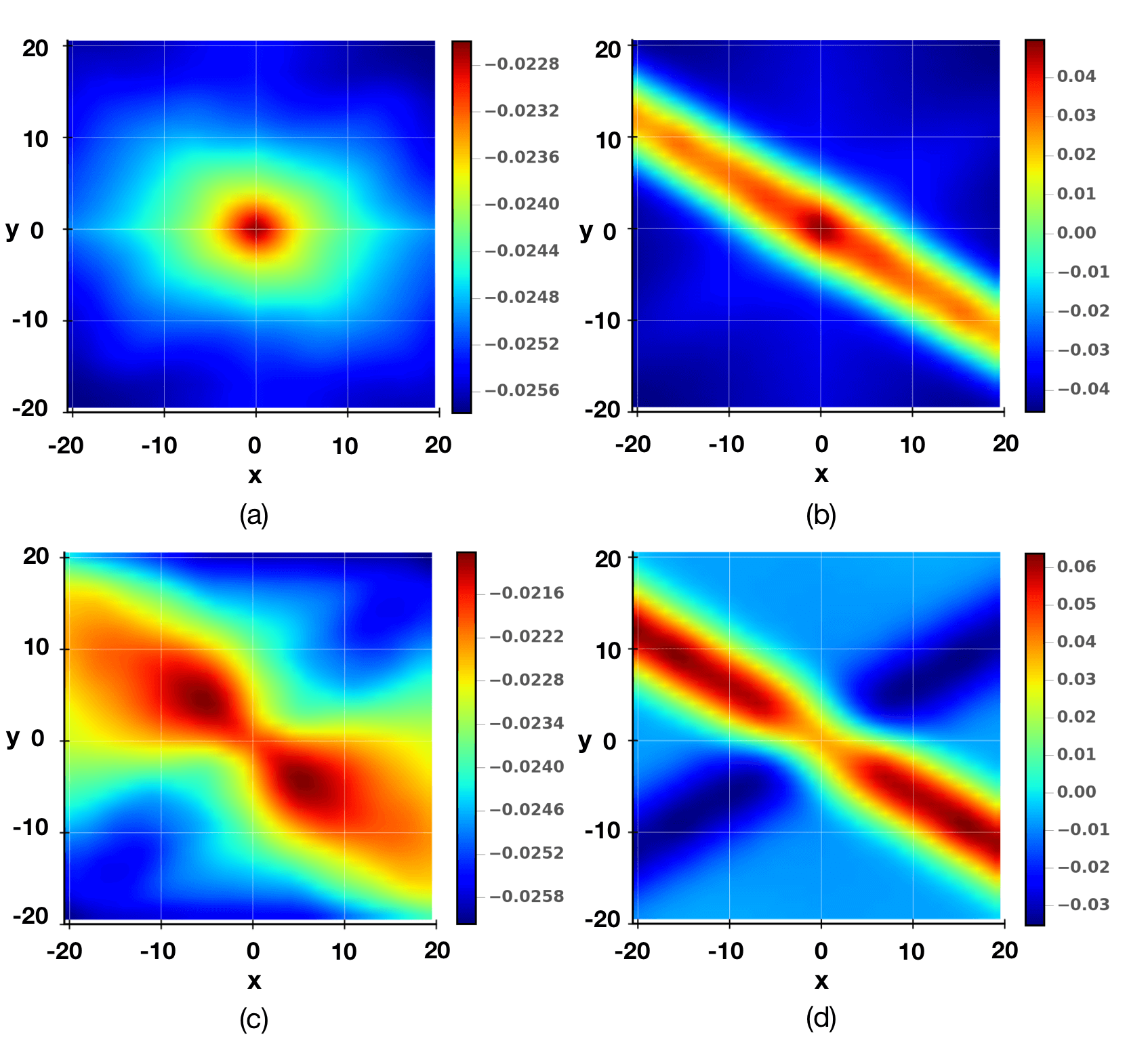} \caption{ {\textbf{First and second principal component of PCA application on thin films of $\boldsymbol{w=0.5}~\boldsymbol{\mu \rm{m}}$ shown in sample coordinates:}} (a) First principal component - Small DIC-test reloaded state strain (0.1 $\%$) (b) First principal component- Large DIC-test reloaded state strain (1 $\%$) (c) Second principal component - Small DIC-test reloaded state strain (0.1 $\%$) (d) Second principal component- Large DIC-test reloaded state strain (1 $\%$). For description of colormaps, see Fig.~\ref{fig:width2}.}
\label{fig:diff1stinvnon}
\end{figure}

\subsection{Distinguishing plasticity regimes for large DIC-testing (1 $\%$) total strain}

The construction of the ``large-reload'' data set was intended to show the delicate handling needed for us to get the cluster separation we require. In particular, when reloading to $1$ $\%$ DIC-testing strain our method does not produce the desired separation. As the DIC-testing strain increases so do strain localization features and shear band sizes. With a shear band spanning the whole specimen, we expect that the statistical correlations differ significantly from the statistical correlations of the ``small-reload'' data set. That is due to the overall effect of localization, from a structural correlation viewpoint: as load is applied to the sample and before strain localization emerges in the form of shear bands, crystal plasticity must be homogeneous and relatively isotropic. This is mainly due to the dislocation sources and obstacles that are placed in an unbiased manner (randomly) across the slip planes in our simulations. However, after localization emerges, correlations are dominated by shear banding and become highly anisotropic. Our methods pick up that transition. Indeed, even in the case of lower reload strain, while some separation is present, for the prior loaded state of 0.1 $\%$ strain and 1 $\%$ strain the distance between clusters is small, while in smaller sized systems (Fig.~\ref{fig:width05}) samples are unclassifiable. Coupled with the invasive nature of the total strain imposed on the DIC-test reloaded state, the inseparability of the clusters makes the samples' prior loaded state history indistinguishable.

Fig.~\ref{fig:inv-noninvcomp} compares the small and largereload DIC-testing regimes. Fig.~\ref{fig:inv-noninvcomp} (b) and (d) show the results of PCA with a clustering algorithm applied to the data set. From Fig.~\ref{fig:inv-noninvcomp} (d) we can see that higher DIC-testing strain renders samples indistinguishable in PCA coordinates. The separation that was present for the low DIC-testing strain (0.1 $\%$) is missing
for higher values. Figs.~\ref{fig:inv-noninvcomp} (a,c) show the strain profiles captured when the sample is reloaded to low (a) DIC-testing strains and high (c).  A more comprehensive comparison for these regimes can be found in Appendix C.3. It is obvious that for higher DIC-testing strain there is much more mixing of the samples, thus the classification algorithm fails. Fig.~\ref{fig:diff1stinvnon} shows another difference between the two DIC testing regines. For small reload strain the first principal component (a) appears relatively isotropic, while it becomes highly anisotropic for larger reload strain (c). This observation extends to other components (e.g., 2nd (b,d)) and is correlated to the emergent anisotropy of strain localization.

\subsection{Dependence of unsupervised learning capacity on pre-processing aspects.}

\begin{figure}[H] \centering
\includegraphics[width=1\textwidth]{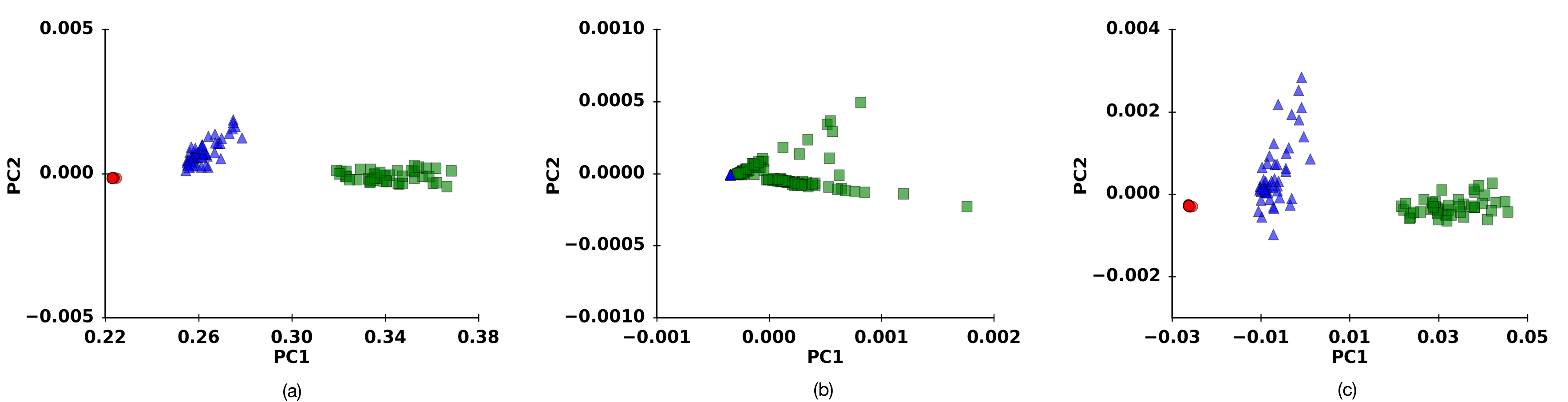}

\caption{ {\textbf{Auto-correlations {\it vs.} cross-correlations for pre-processing -- Example of PCA projection maps for $\boldsymbol{w=2}~\boldsymbol{\mu \rm{m}}$:}} The colors follow the definition of Fig.~\ref{fig:width2}. $w=2 \micron$. (a) $\omega_1$,$\omega_1$ auto-correlations (b) $\omega_1$,$\omega_2$ cross-correlations (c) $\omega_2$,$\omega_2$ auto-correlations.}
\label{fig:w2icorr}
\end{figure}

As discussed in Sec.~2.3, the discretization scheme defines the form and dimensions of the correlation functions to which we apply a PCA transformation. We can choose to examine correlations between different local states $\omega$. We can categorize samples based on their deformation history either for $\omega_1$,$\omega_1$ auto-correlations or $\omega_2$,$\omega_2$ auto-correlations. We find that cross-correlations aren't helpful for classifying samples according to their deformation levels. Figs.~\ref{fig:w2icorr} shows results obtained from various correlation functions: in general, as $w$ decreases, we observe that distances between each cluster are also decreasing. In particular, Fig.~\ref{fig:w2icorr} (b) shows that the distances in each cluster are measured in an order of magnitude $10^{-4}$ to $10^{-3}$ while in Fig.~\ref{fig:w2icorr} (a),(c) the order of magnitude is $10^{-2}$ to $10^{-1}$. This difference in Fig.~\ref{fig:w2icorr} (a),(c) is enough for the clustering algorithm to find the different deformation levels and classify with 100 $\%$ accuracy our data set.

\begin{figure}[H] \centering
\includegraphics[width=0.72\textwidth]{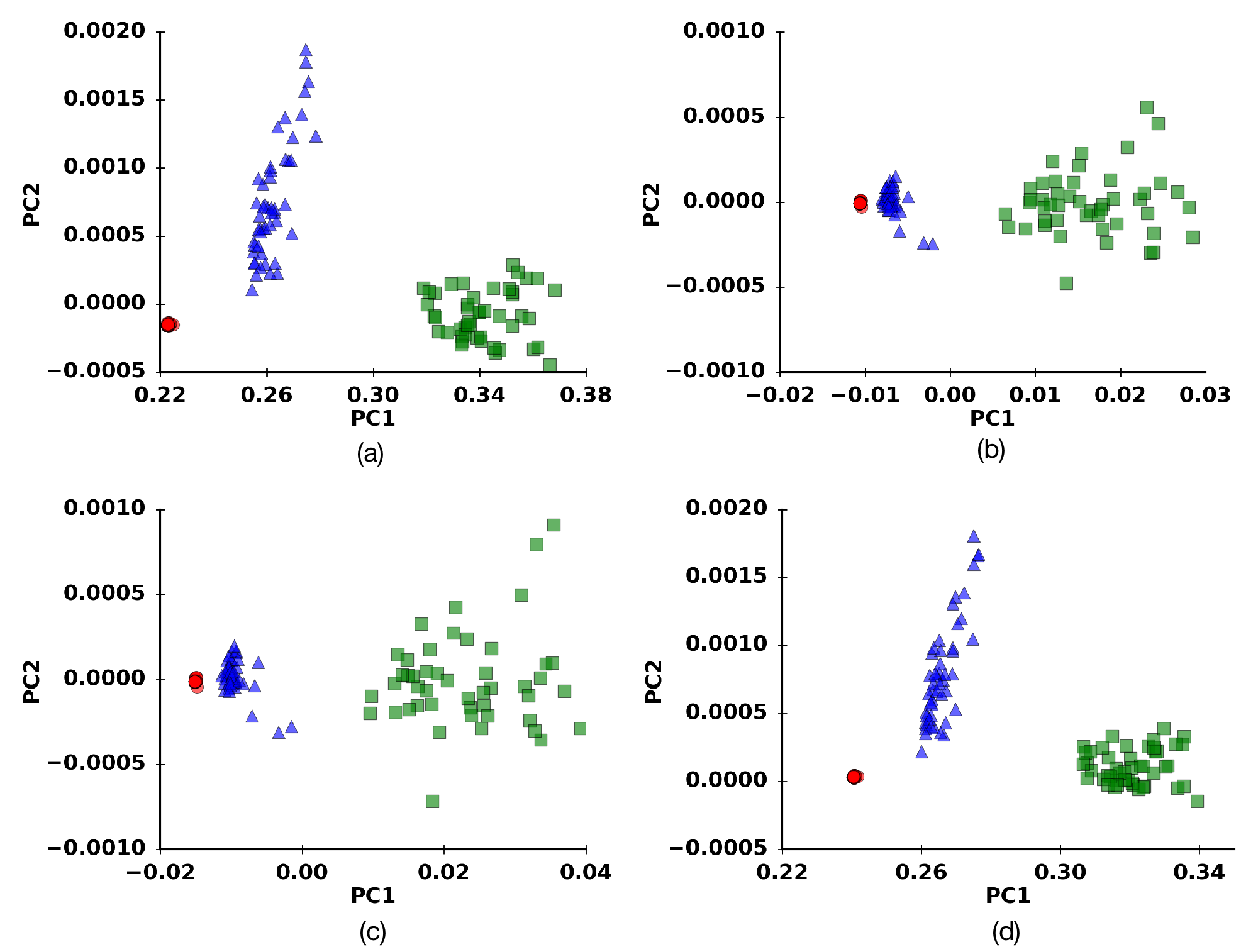}

\caption{ {\textbf{Effect of strain invariant type for pre-processing -- Examples of PCA projection maps:}} The colors follow the definition of Fig.~\ref{fig:width2}. (a) $\omega_1$,$\omega_1$ auto-correlations. $w=2 \micron$. Isotropic measurements. (b) $\omega_1$,$\omega_1$ auto-correlations. $w=2 \micron$. $J_2$ invariant. (c) $\omega_2$,$\omega_2$ auto-correlations. $w=1 \micron$. Isotropic measurements. (d) $\omega_2$,$\omega_2$ auto-correlations. $w=1 \micron$. $J_2$ invariant. } 
\label{fig:j2inv}
\end{figure}

Another choice we can make is the quantity that characterizes the
microstructure. Until now, we considered an isotropic measure of the
total deformation strain in the sample. Our classification scheme
produces similar results if we use the more common 2nd invariant of
the strain deformation tensor, $J_2 = \eps_{ik}
\eps_{ki}$. Fig.~\ref{fig:j2inv} shows the results for different
microstructural measure calculations. For larger systems ($w=1$,$w=2$)
the only notable difference is the overall variance of the data in PCA
coordinates.

\begin{figure}[H] \centering
\includegraphics[width=0.8\textwidth]{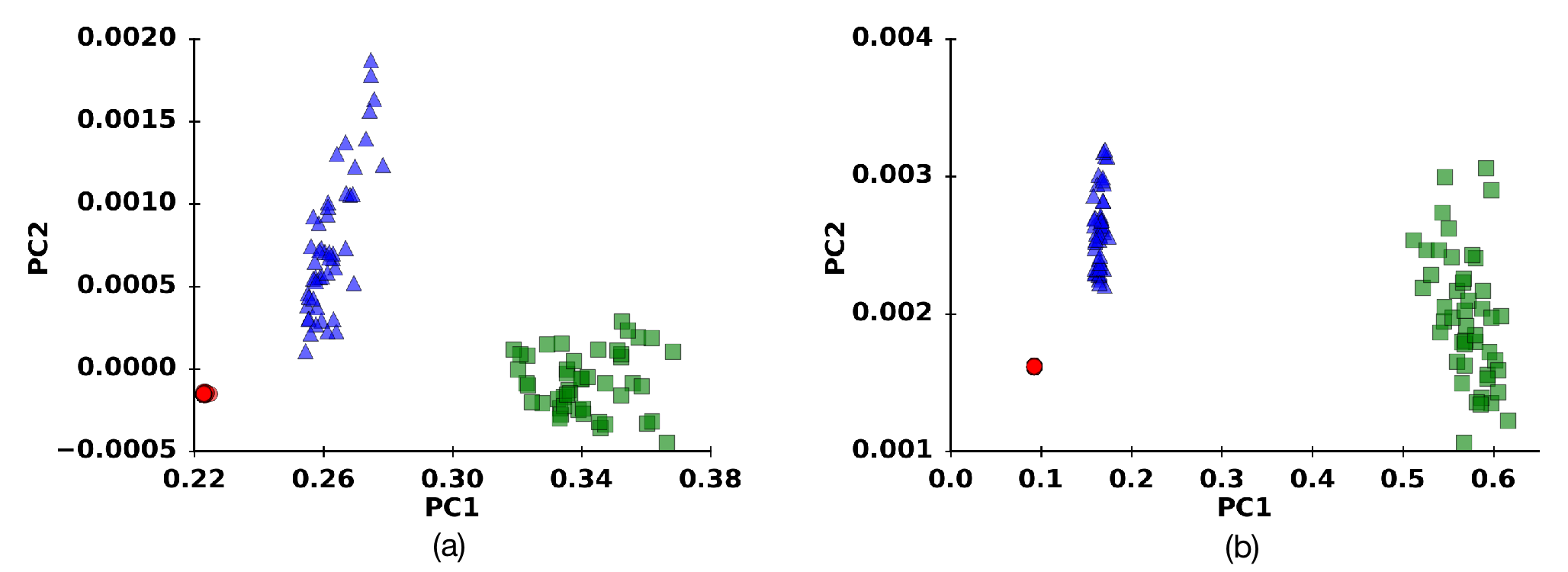}
\caption{ {\textbf{Total {\it vs.} Residual/Plastic strain for pre-processing: Examples of PCA projection maps:}} The colors follow the definition of Fig.~\ref{fig:width2}. (a) Plastic strain. $\omega_1$,$\omega_1$ auto-correlations. $w=2 \micron$ (b) Total strain. $\omega_1$,$\omega_1$ auto-correlations. $w=2 \micron$}
\label{fig:resid_plast}
\end{figure}

Finally, we may use the plastic strain as the microstructural deformation state variable (following Kalidindi -it is not actually a thermodynamic state variable), which would effectively correspond to applying DIC on the test unloaded state dislocation ensembles. The test unloaded state would be the state that would emerge after unloading the sample on the test reloaded state. Fig.~\ref{fig:resid_plast} shows that classification still works and there is an observable difference of the data variance in PCA coordinates. For more figures on the differences in pre-processing aspects see Appendix C.4.

\newpage
\subsection{Independence from the choice of discretization schemes and dimension reduction methods}

\begin{figure}[H] \centering
\includegraphics[width=0.95\textwidth]{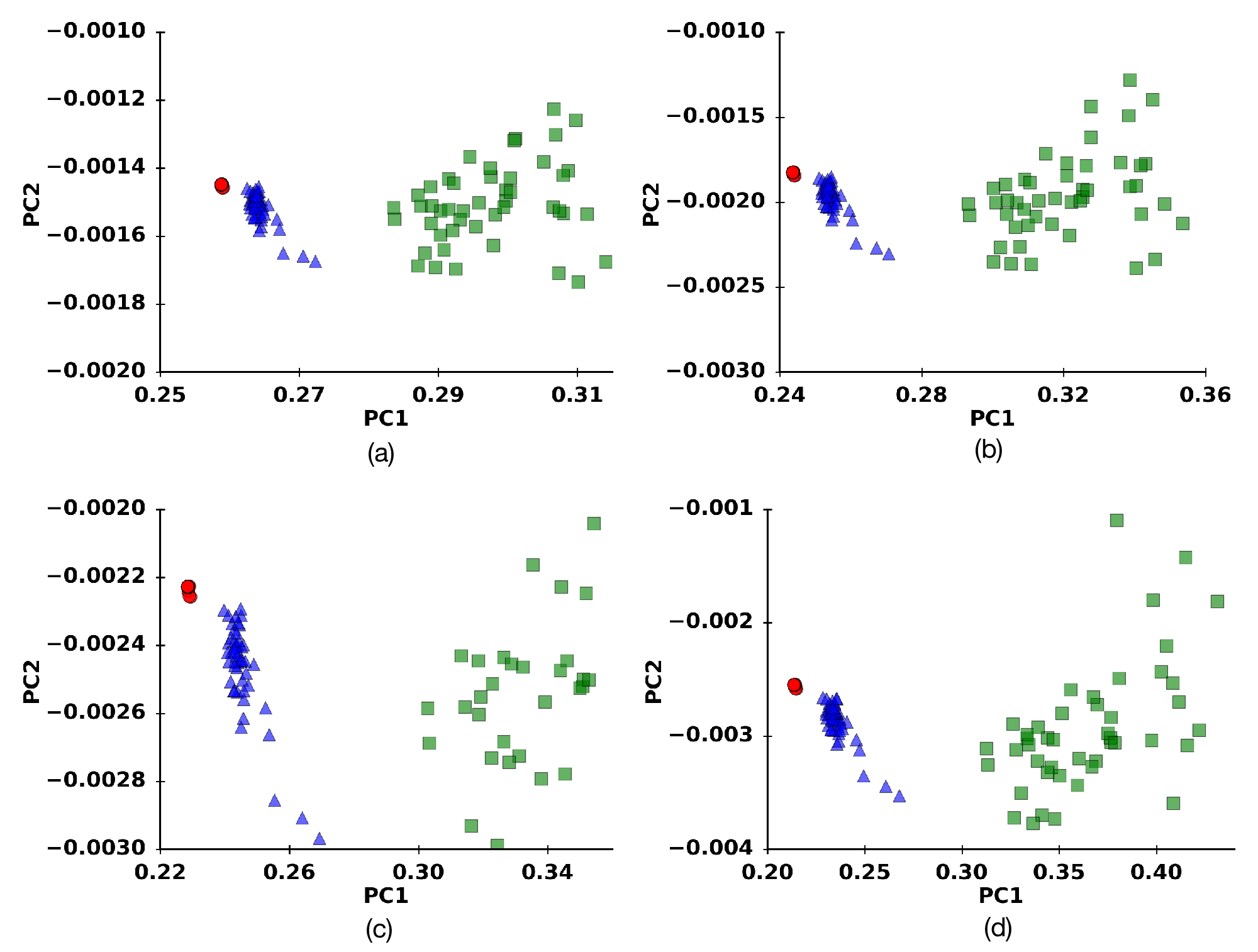}
\caption{ {\textbf{Effect of discretization schemes on pre-processing: Examples of PCA projection maps for $\boldsymbol{w=1}~\boldsymbol{\mu \rm{m}}$:}} The colors follow the definition of Fig.~\ref{fig:width2}. $\omega_1$,$\omega_1$ auto-correlations. (a) 2 local states (b) 3 local states (c) 4 local states (d) 5 local states}
\label{fig:localstates}
\end{figure}

We find that our methods are not sensitive to reasonable variations of the microstructural binning of the local strain variable. As a test, we discretize the microstructure into $L=2$, 3, 4 and 5 parts. We are able to distinguish the initial deformation history of all the samples when calculating the $\omega_1$,$\omega_1$ auto-correlations and the $\omega_2$,$\omega_2$ auto-correlations. These results are independent of the discretization scheme (i.e. the number of local states used). Fig.~\ref{fig:localstates} shows the results for data samples of $w=1$ $\micron$, as the number of local states $L$ increases. Clustering and classification is possible, and the clustering algorithm has 100 $\%$ accuracy independently of the number of local states, but the overall noise of the data increases with the number of local states. 
{
The noise is due to the use of a fixed number of DDD simulations for each prior strain level. The signal strength in each correlation function increases with system size and the number of dislocations, but decreases as the data is distributed into more bins $L$. This effect is more pronounced for cross-correlations because they decrease for short distances and our correlation function range is truncated. Hence we do not obtain classifiable results for any cross correlations.
}

\newpage
While PCA is one of the most common and useful tools for dimensionality reduction, some data sets could be so large that it is impractical.  With that in mind, we compared our PCA results with other common algorithms, such as Incremental Principal Component Analysis (IPCA) and the Truncated Singular Value Decomposition (TSVD). IPCA uses a different form of processing a data set that allows for partial computations which in most cases match the results of PCA. Incremental PCA stores estimates of component variances and updates the variance ratio of a component incrementally. It is faster and uses memory more efficiently than PCA. TSVD on the other hand, implements a variant of singular value decomposition (SVD) that only computes the largest singular values. Given that PCA works on the basis of the singular value decomposition, we expect little to no difference with this method.

\begin{figure}[H] \centering
\includegraphics[width=\textwidth]{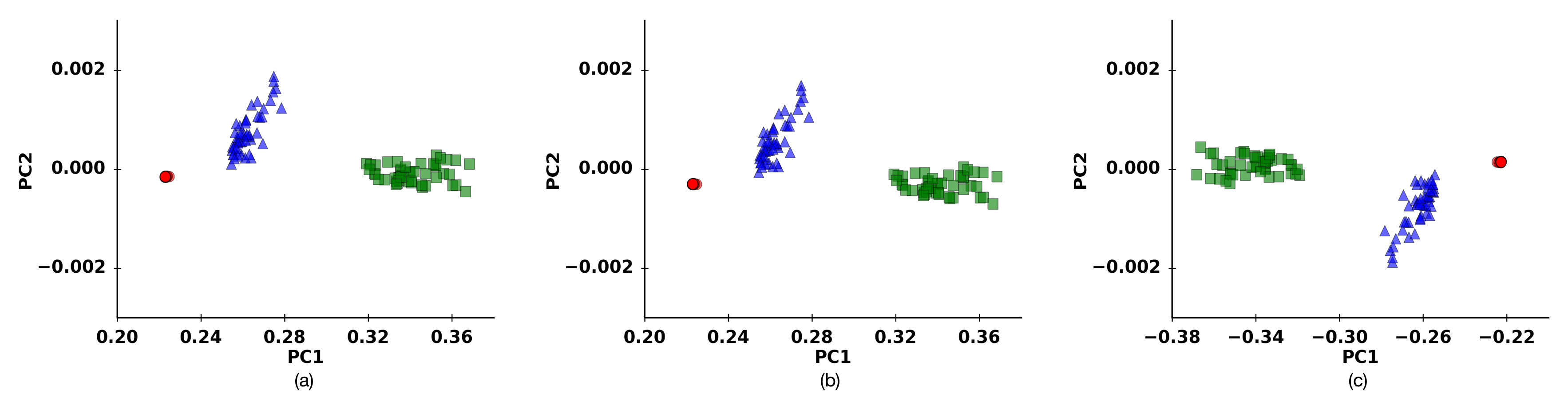}
\caption{ {\textbf{Comparison of different dimensionality reduction methods for $\boldsymbol{w=2}~\boldsymbol{\mu \rm{m}}$:}} The colors follow the definition of Fig.~\ref{fig:width2}. $\omega_1,\omega_1$ auto- correlations. (a) PCA (b) TSVD (c) IPCA}
\label{fig:diffpca}
\end{figure}

No significant differences are seen when applying some of these variations of PCA to our data sets. The  results are shown on Fig.~\ref{fig:diffpca}. Note that no additional parameters, other than the initialization of the different methods, have been modified; in particular, the same clustering algorithm is used as with the PCA methods. The TSVD results do not display any differences from regular PCA, besides slight changes in data variance and data cluster positions. The IPCA results, on the other hand, are mirrored from the PCA results in both the PC1 and PC2 axes (negative values). If we calculated the absolute values we would see just minor differences in data variance and cluster positions as in the TSVD results. 

\section{Summary}

We examined the applicability of ML algorithms to practical and relatively inexpensive experimental methods for the detection of mechanical deformation and crystal plasticity, such as DIC. The principal question was whether it is possible to detect the degree of {\em prior} plastic deformation of thin films, especially when they display significant plasticity size effects. Our  conclusion is that ML algorithms can achieve our objective with varying levels of success in various situations. Through mimicking the DIC protocol in two dimensional discrete dislocation plasticity simulations, we identified realistic cases (single and double slip thin films with widths larger than 1 $\micron$) where data clustering and classification is possible, based on the degree of prior plastic deformation. However, when size effects come into play, we found that such clustering and classification becomes gradually more difficult, since the intrinsic, plasticity induced crackling noise causes large variance in smaller systems. Our simulations uncovered a crucial parameter for the applicability of our methods, namely the DIC-testing total strain during reloading. While for a small-reload level of 0.1 $\%$ (half of the commonly defined engineering yield stress, found at at engineering strain 0.2 $\%$), our methods are highly successful, they are clearly not successful one order of magnitude higher, at 1 $\%$.  In summary, we demonstrated that the detection of prior plastic deformation is possible through the use of principal component analysis and the CkNN algorithm, for films wider than 0.5 $\micron$ and DIC-testing total strains smaller than 0.5 $\%$. Our future goals include quantitatively predicting the prior strain (and its uncertainty) from the test strain, and also applications to actual DIC data from experiments.

The main difference between our simulation and actual DIC data sets is the resolution. Since we obtain direct strain measurements we assume we have infinite resolution for examining our data set, in contrast to DIC, which has limited resolution. DIC is limited in that regard in a 3-fold way: the average observable distance in experiments, the smallest  distance between DIC markers and the resolution of the machinery that is used. 

\section{Acknowledgements} We would like to thank D. Wheeler and S. Kalinindi for illuminating discussions.  We would like to thank Marilyn Y. Vasquez Landrove for sharing the CkNN clustering code \citep{berry2017}, and Erik Van Der Giessen for insightful comments on our work. We also acknowledge funding from Department of Commerce under award No. 1007294R (SP). 

\newpage
%\section*{References}

\newpage
\appendix

\section{Discrete Dislocation Dynamics}

\begin{figure}[tbh] \centering
\includegraphics[width=0.5\textwidth]{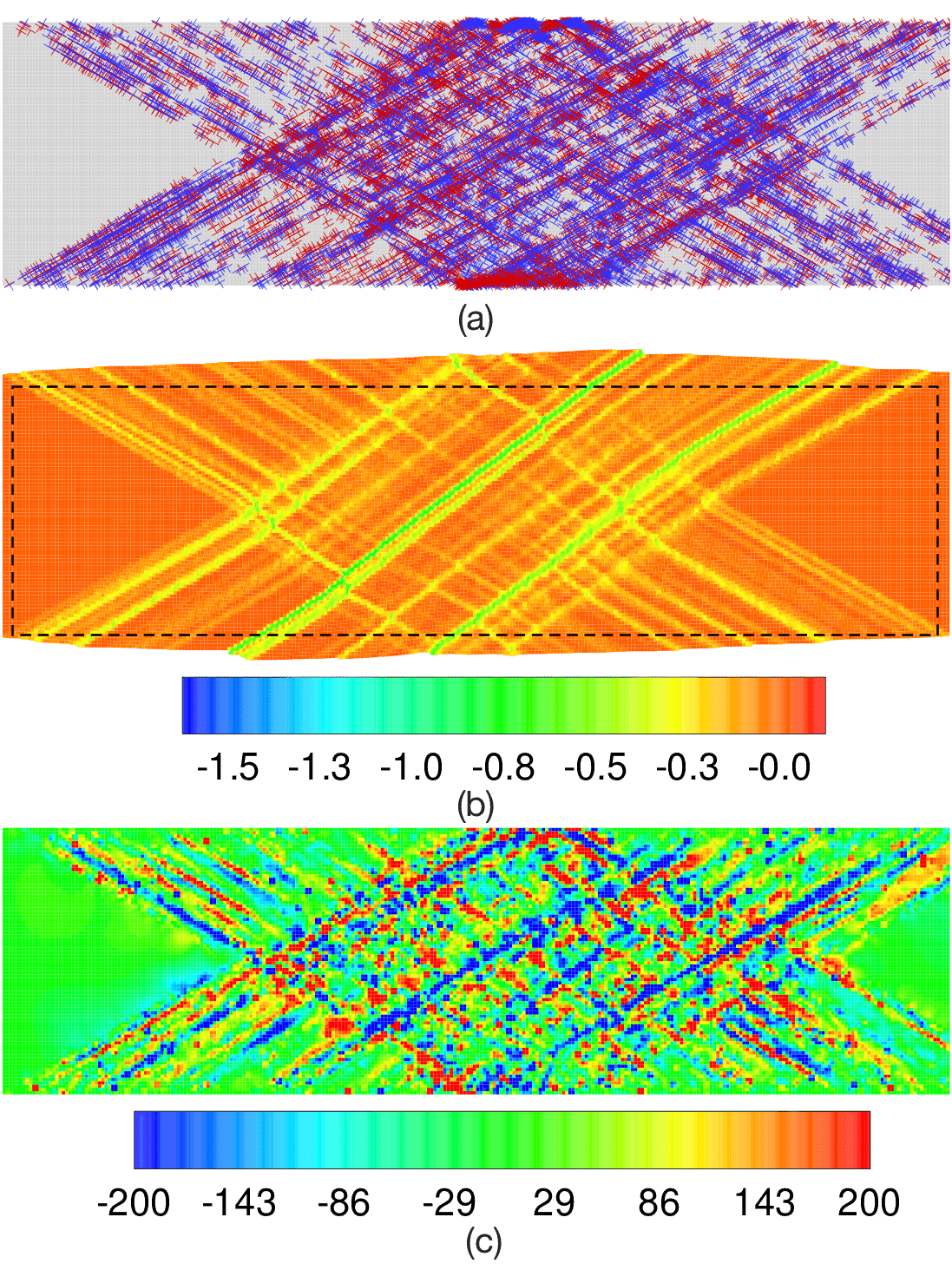}
\caption{\textbf{Boundary plastic steps and stress/strain profiles in dislocation configurations:} (a) A dislocation snapshot at 10 $\%$ strain for a sample of width w=2 and double slip systems. A high density of dislocation is observed, close to $10^{15}/m^2$. T's and $\bot$'s have been used to describe the dislocations that create positive stress (T's) and negative stress ($\bot$'s) on the different slip systems. (b) The $\epsilon_{xx}$-strain profile that corresponds to the dislocation configuration of (a) is shown in the displaced coordinates. The boundary is filled with rough plastic steps.  The dashed rectangular box denotes the region for which we consider spatial strain correlations, that are then used for ML purposes. The strain map is unitless. (c) The shear stress $\tau_{xy}$ of the configuration in (a,b) is shown. The stress map is in MPa.}
\label{fig:intro3}
\end{figure}

In our DDD simulation, plastic flow occurs by the nucleation and glide
of edge dislocations, on single or double slip systems. With the typical
Burgers vector of FCC crystals being $b=0.25 \rm{~nm}$, we study sample
widths ranging in powers of 2 from $w=0.125 \micron$ to $2 \micron$ with
$\alpha=h/w=4$--$32$. The lateral edges ($x=0,w$) are traction free,
allowing dislocations to exit the sample. Loading is taken to be
ideally strain-controlled, by prescribing the $y$-displacement at the
top and bottom edges ($y=0,h$). The applied strain rate (for both
loading and unloading regimes), $\dot{h}/h=10^4 \text{s}^{-1}$, is
held constant across all our simulations, similar to experimental
practice.  Plastic deformation of the crystalline samples is described
using the discrete dislocation framework for small
strains~\citep{vandergiessen1995}, where the determination of the
state in the material employs superposition.  As each dislocation is
treated as a singularity in a linear elastic background solid with
Young's modulus $E$ and Poisson ratio $\nu$, whose analytic solution
is known at any position, and because the sample does not extend to
infinity: the displacement, strain and stress fields need to be
corrected by smooth image fields (denoted by $\hat{\ }$ below) to satisfy boundary conditions at the edges. Hence, the displacements $u_i$, strains $\varepsilon_{ij}$, and stresses $\sigma_{ij}$ are written as

%this field needs to be corrected by a smooth image field $(\hat{\ })$ to ensure that actual finite volume boundary conditions are satisfied.  

\begin{equation}
\label{eq:superposition} u_i = \tilde{u_i}+\hat{u_i}, \; \varepsilon_{ij} = \tilde{\varepsilon}_{ij}+\hat{\varepsilon}_{ij}, \; \sigma_{ij} = \tilde{\sigma}_{ij}+\hat{\sigma}_{ij}, \end{equation} where the ($\tilde{\ }$) field is the sum of the fields of all $N$ dislocations in their current positions, i.e. 

\begin{equation}
\label{eq:dislocation-field}
\tilde{u}_i=\sum_{J=1}^{N}\tilde{u}_i^{(J)}, \; \tilde{\varepsilon}_{ij}=\sum_{J=1}^{N}\tilde{\varepsilon}_{ij}^{(J)}, \; \tilde{\sigma}_{ij}=\sum_{J=1}^{N}\tilde{\sigma}_{ij}^{(J)}.
\end{equation} 
Image fields are obtained by solving a linear elastic boundary value problem using finite elements with the boundary conditions changing as the dislocation structure evolves under the application of mechanical load. At the beginning of the calculation, the crystal is stress free and there are  no mobile dislocations. This corresponds to a well-annealed sample,  yet with pinned dislocation segments left that can act either as dislocation sources or as obstacles. Dislocations are generated from sources when the resolved shear stress $\tau$ at the source location is sufficiently high ($\tau_{\rm nuc}$) for a sufficiently long time ($t_{\rm nuc}$). We consider bulk dislocation sources and obstacles~\citep{vandergiessen1995}. A dislocation configuration of one of the simulations, at $10\%$ total strain, is shown in Fig.~\ref{fig:intro3}(a), together with the $xx-$component of the total strain (b) and the shear stress (c).

The bulk sources are randomly distributed over slip planes at a density $\rho^{\rm{bulk}}_{\rm{nuc}}= 60 \micron^{-2}$, while their strength is selected randomly from a Gaussian distribution with mean value $\bar{\tau}_{\rm nuc} = 50$ MPa and 10\% standard deviation. Bulk sources are designed to mimic the Frank-Read mechanism in two dimensions~\citep{hirth1982}, such that they generate a dipole of dislocations at distance $L_{\rm nuc}$, when activated. The initial distance between the two dislocations in the dipole is 

\begin{equation}
\label{eq:L_nuc} L_{\rm nuc}=
\frac{E}{4\pi(1-\nu^2)}\frac{b}{\tau_{\rm nuc}},
\end{equation} 
at which the shear stress of one dislocation acting on the other is
balanced by the local shear stress. Surface dislocation sources are
successively placed at opposite ends of slip planes, which corresponds
to a surface density of around $\rho^{sur}_{\rm{nuc}}=175/
\micron$. Once a single dislocation is generated from a surface source, it is put at $10b$ from the free surface. Under this circumstance, our surface nucleated dislocation has an effective nucleation strength of 312 MPa~\citep{papanikolaou2017}. We only consider glide of dislocations, neglecting the possibility of climb. The evolution of dislocations is determined by the component of the Peach-Koehler force in the slip direction. For the $I$th dislocation, this is given by 

\begin{equation}
\label{eq:P-K} 
f^{(I)} = \boldsymbol{n}^{(I)}\cdot\left(\boldsymbol{\hat{\sigma}}+\sum_{J\neq I}{\boldsymbol{\tilde{\sigma}}}^{(J)}\right)\cdot\boldsymbol{b}^{(I)}, 
\end{equation} 
where $\boldsymbol{n}^{(I)}$ is the slip plane normal and $\boldsymbol{b}^{(I)}$ is the Burgers vector of dislocation $I$. This force will cause the dislocation $I$ to glide, following over-damped dynamics, with velocity 

\begin{equation}
\label{eq:B-v} v^{(I)} = \frac{f^{(I)}}{B},
\end{equation} 
where $B$ is the drag coefficient. In this paper, its value is taken
as $B=10^{-4}$Pa s, which is representative for aluminum. Each sample
contains a random distribution of forest dislocation obstacles and
surface dislocation sources, as well as a random distribution of bulk
dislocation sources.  Once nucleated, dislocations can either exit the
sample through the traction-free sides, annihilate with a dislocation
of opposite sign when their mutual distance is less than $6b$, or
become pinned at an obstacle. Point obstacles are included to account
for the effect of blocked slip caused by precipitates and forest
dislocations on out-of-plane slip systems that are not explicitly
described. They are randomly distributed over slip planes with a
constant density that corresponds on average, to one source, either
surface or bulk, for every 8 randomly-distributed obstacles. In this way the densities of sources and obstacles remains the same as the sample dimensions change, but there is a statistical preference towards always accompanying sources with obstacles in order to avoid statistical outlier behaviors. We model obstacles in a simple way where a dislocation stays stays pinned until its Peach-Koehler force exceeds the obstacle-dependent value $\tau_{\rm obs}b$. The strength of the obstacles $\tau_{\rm obs}$ is taken to be $300$ MPa with 20 $\%$ standard deviation. Our simulations are carried out for material parameters that are reminiscent of aluminum: $E = 70$ GPa,$\nu = 0.33$. We consider $50$ random realizations for each parameter case. 

The simulation is carried out in an incremental manner, using a time step that is a factor 20 smaller than the nucleation time $t_{\rm nuc}=10\:$ns. At the beginning of every time increment, nucleation, annihilation, pinning at and release from obstacle sites are evaluated. After updating the dislocation structure, the new stress field in the sample is determined, using the finite element method to solve for the image fields~\citep{vandergiessen1995}. 

\section{Microstructural discretization and $N$-point correlation statistics}

To avoid confusion, we will be referring to the strain fields as the ``microstructure'', following Kalidindi's terminology~\citep{kalidindi2012}. However, it is clear that the complete microstructure should also include the material and all of its properties that go into determining the strain field.  The discretization of our microstructure is the separation of the local strain information into various stages (low,intermediate and high local strains). The continuous local state variable h, the local state space $\cal{H}$ and the microstructure function $m(h,x)$ are used to represent a {\em single} microstructure in a digital format. The local state space $\cal{H}$ can be thought of as the complete collection of all the necessary state variables that are needed to uniquely define the material structure at a given location. The local state variable $h$ is one point in the local state space, or one configuration of state variables. In
our simulations, we consider the isotropic strain measure $\phi$ which takes the role of the continuous local state variable $\phi\equiv h$. 
The microstructure $\mu(x)$ can be described by a distribution $m(h,x)$, the microstructure function, for a  local state $h$ at a position $x$:
\begin{equation}
\mu(x)=\int_{H}^{ }hm(h,x)dh \;.
\end{equation}

The domain of a microstructure is discretized in both real and state space. Binning in real space follows naturally from the DDD simulation method, dividing the total volume into $S$ rectangular bins, neglecting bins that are too close to the boundary (see Appendix A-- Fig.~\ref{fig:intro3}). The local state space $\cal{H}$ can be binned by expanding $m$ in a set of basis functions.
We use the primitive basis function from PyMKS, $\Lambda_l(h)$, which has a triangular form 
\begin{equation}
\Lambda_{l}(h)={\rm max}(1-\left | \frac{h-h_l}{\delta h} \right |,0)
\end{equation}
where  $\delta h$ is the bin width. For linear binning of $H$ into $L-1$ bins, $\delta h=H/(L-1)$,   with any bin defined by the edges $l\in[1,L]$ with values $h_{l-1}$, $h_l$. 
The discretized microstructure function is then
\begin{equation}
m[l,s]  = \frac{1}{\Delta x}\int_{H}^{}\int_{s}^{ }\Lambda_l(h)m(h,x)dxdh
\end{equation}
in spatial bin $s$ and state bin $l$.
In the above expressions, variables in round brackets are continuous
and variables in square brackets are discrete. 

A microstructure function discretized with this basis is subject to the constraint 
\begin{equation}
\sum_{l=1}^{L}m[l,s]=1
\end{equation}
which is equivalent to saying that every location (spatial cell) is filled with some configuration of local states. 

N-point spatial correlations provide a way to quantify material structure, using statistics. 1-point statistics is based on the probability that a specified local state will be found in any randomly selected spatial bin in a microstructure~\citep{niezqoda2008,fullwood2009,niezqoda2010}. 1-point statistics computes the volume fractions of the local states in the microstructure.

\begin{equation}
f[l]=\frac{1}{S}\sum_{s}m[s,l]
\end{equation}

In this equation, $f[l]$ is the probability of finding the local state $l$ in any randomly selected spatial bin in the microstructure,
$m[s,l]$ is the microstructure function (the digital representation of the microstructure), $S$ is the total number of spatial cells in the microstructure and $s$ is a specific spatial cell.

2-point spatial correlations (also known as 2-point statistics) contain information about the fractions of local states as well as the first order information on how the different local states are distributed in the microstructure. 2-point statistics is based upon the probability of having a vector placed randomly in the microstructure with one end of the vector be on one specified local state and the other end on another specified local state. This vector could have any length or orientation that the discrete microstructure allows. The equation for 2-point statistics reads:

\begin{equation}
f[r|l,l']=\frac{1}{S}\sum_{s}^{ }m[s,l]m[s+r,l']
\label{2point}
\end{equation}

In this equation $f[r|l,l']$ is the conditional probability of finding the local states $l$ and $l'$ at a distance and orientation away  from each other defined by the vector $r$. When the 2 local states are the same $l=l'$, the correlation is called an auto-correlation. If the 2 local states are not the same, it is a cross-correlation. Our algorithm combines 2-point spatial correlations and the primitive basis function. Given our microstructure and through simple one line commands of PyMKS we find the correlations we want in the spatial bins. The relevant information of auto and cross-correlations are stored into matrices for further processing. The discretization of the microstructure into the arbitrary chosen local states allows for multiple cross and auto-correlations. 

In our case we discretize our microstructure in 3 different bins, with 3 possible local states, at low, intermediate and high local strains. These local states create the following possible correlations: let $\omega_1$ be local state 1, $\omega_2$ local state 2 and $\omega_3$ local state 3. We have ($\omega_1$,$\omega_1$) auto-correlations, ($\omega_2$,$\omega_2$) and ($\omega_3$,$\omega_3$) auto-correlations. Furthermore, it is also possible to observe ($\omega_1$,$\omega_2$), ($\omega_1$,$\omega_3$) and ($\omega_2$,$\omega_3$)
cross-correlations.

\newpage
\section{Results}

\subsection{Distinguishing plasticity regimes for small DIC-testing strain (0.1 $\%$)}

\begin{figure}[H] \centering
\includegraphics[width=1\textwidth]{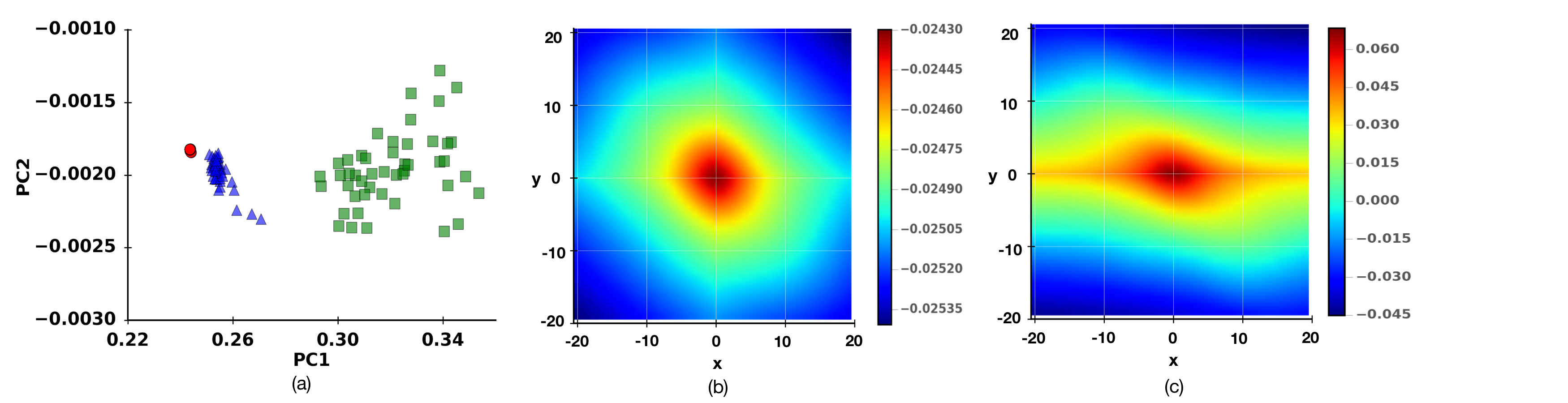}
\caption{ {\textbf{ $\boldsymbol{w=1}~\boldsymbol{\mu \rm{m}}$ -- 2D projection of PCA results for thin films -- Double slip system:}} $\omega_1$,$\omega_1$ auto-correlation. The colors follow the definition of Fig.~\ref{fig:width2}. (a) Projection of data set on first two principal components. (b) First principal component of PCA, shown in sample coordinates (Fig.~\ref{fig:schematic}). (c) Second principal component of PCA, shown in sample coordinates (Fig.~\ref{fig:schematic}). For description of colormaps, see Fig.~\ref{fig:width2}.}
\label{fig:width1}
\end{figure}

\begin{figure}[H] \centering
\includegraphics[width=0.72\textwidth]{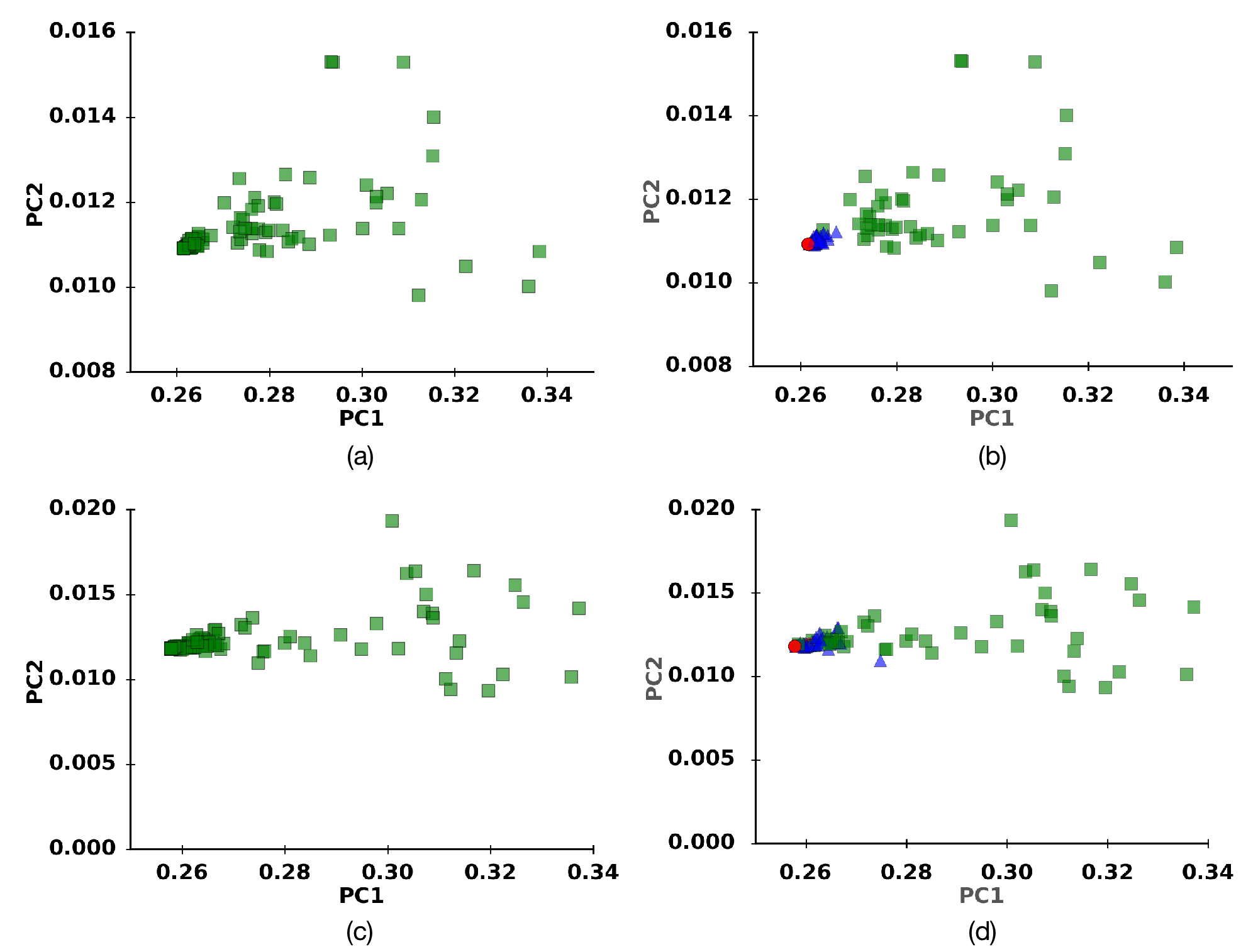}
\caption{ {\textbf{$\boldsymbol{w=0.25}~\boldsymbol{\mu \rm{m}}$, $\boldsymbol{w=0.125}~\boldsymbol{\mu \rm{m}}$ -- 2D projection of PCA results for thin films -- Double slip system :}} ($\omega_1$,$\omega_1$) auto-correlations. The colors follow the definition of Fig.~\ref{fig:width2}. (a) Projection of data set on first two principal components with a clustering algorithm applied to the data set. $w=0.25 \micron$. (b) Projection of data set on first two principal components without a clustering algorithm applied to the data set. $w=0.25 \micron$. Actual representation of initial deformations. (c) Projection of data set on first two principal components with a clustering algorithm applied to the data set. $w=0.125 \micron$. (d) Projection of data set on first two principal components without a clustering algorithm applied to the data set. Actual representation of initial deformations. $w=0.125 \micron$.}
\label{fig:2dpca}
\end{figure}

This section of the Appendix contains results that correspond to Section 3.3 of the main text. Figs.~\ref{fig:width1} and ~\ref{fig:2dpca} show results for samples with $w=1$, $w=0.25$ and $w=0.125$ $\micron$. In Fig.~\ref{fig:2dpca} (b) and (d) we show how a correct classification of the samples should look like, in contrast to (a) and (c), which show the results of our classification scheme. 

\subsection{Distinguishing plasticity regimes for single slip samples}

This section contains results from simulations of single slip systems. In particular, we show the results for $w=0.25$ and $w=1$ $\micron$ to emphasize the likeness between results of single slip and double slip systems. 

\begin{figure}[H] \centering
\includegraphics[width=0.72\textwidth]{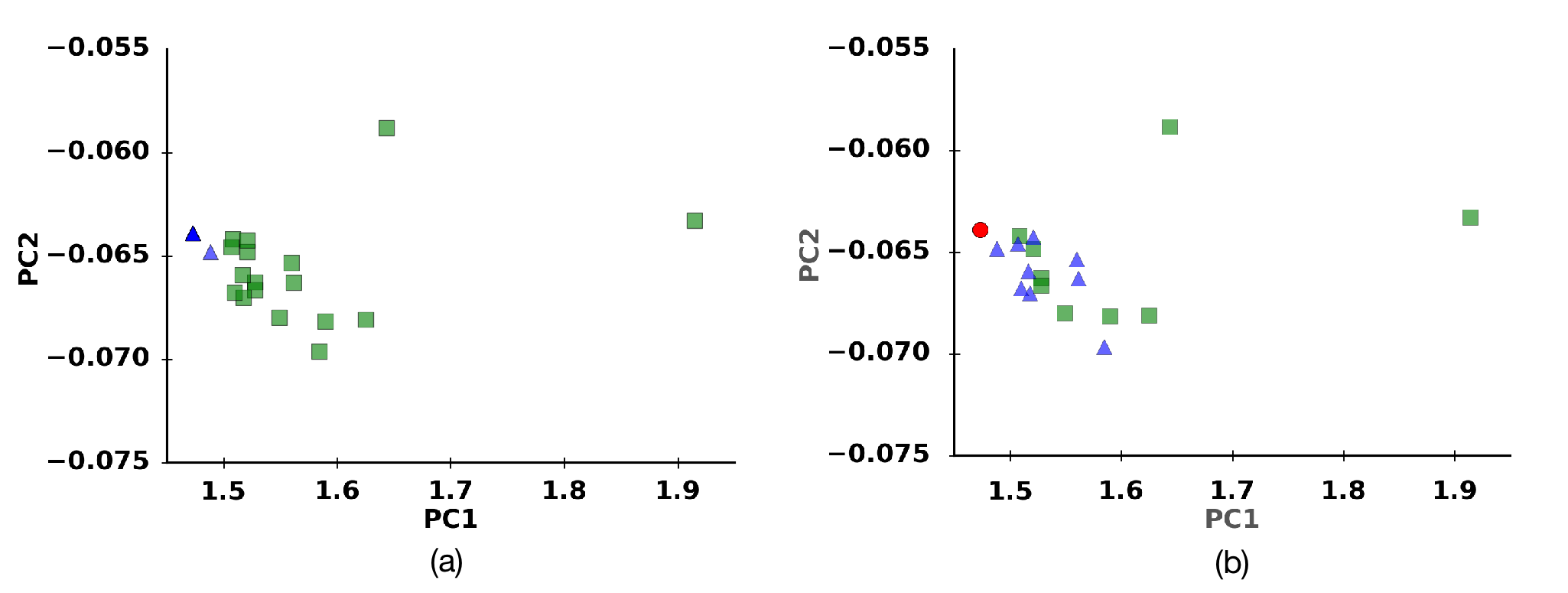}
\caption{ {\textbf{$\boldsymbol{w=0.25}~\boldsymbol{\mu \rm{m}}$ -- 2D projection of PCA results for thin films -- Single slip system:}} $\omega_1$,$\omega_1$ auto-correlation. The colors follow the definition of Fig.~\ref{fig:width2}. (a) Projection of data set on first two principal components with a clustering applied to the data set, demonstrating failure in clustering. (b) Projection of data set on first two principal components without a clustering algorithm applied to the data set, justifying the failure of clustering in (a).}
\label{fig:025oneslip}
\end{figure}

\begin{figure}[H] \centering
\includegraphics[width=0.52\textwidth]{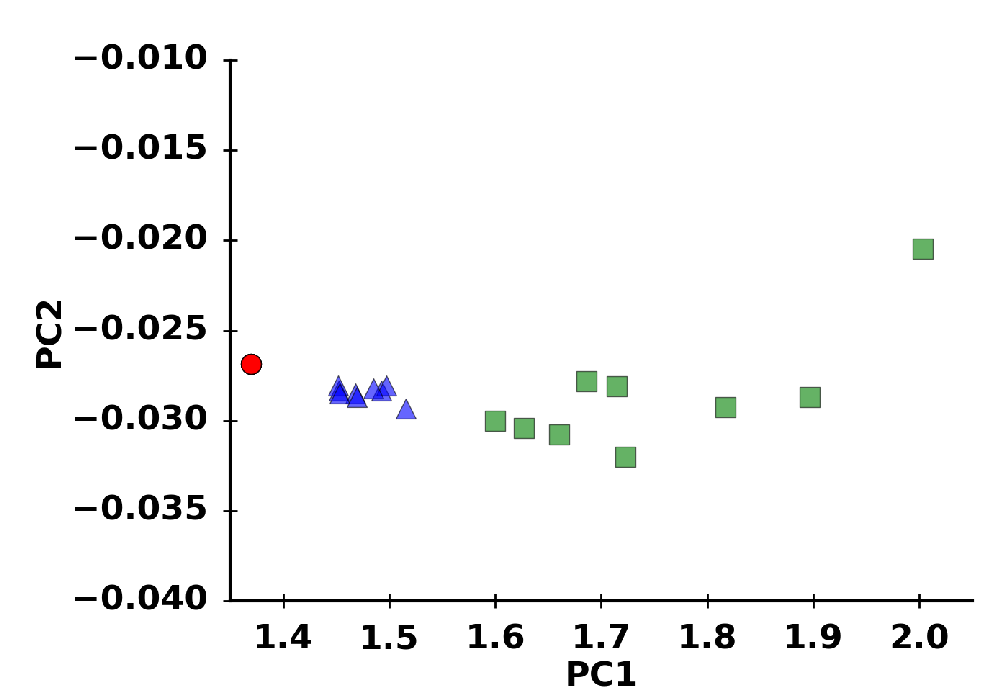}
\caption{ {\textbf{$\boldsymbol{w=1}~\boldsymbol{\mu \rm{m}}$ -- 2D projection of PCA results for thin films -- Single slip system:}} $\omega_1$,$\omega_1$ auto-correlation. The colors follow the definition of Fig.~\ref{fig:width2}. Projection of data set on first two principal components with a clustering applied to the data set, demonstrating failure in clustering.}
\label{fig:oneslip_1}
\end{figure}

\newpage
\subsection{Distinguishing plasticity regimes for large DIC-testing strain (1 $\%$)}

\begin{figure}[H] \centering
\includegraphics[width=0.68\textwidth]{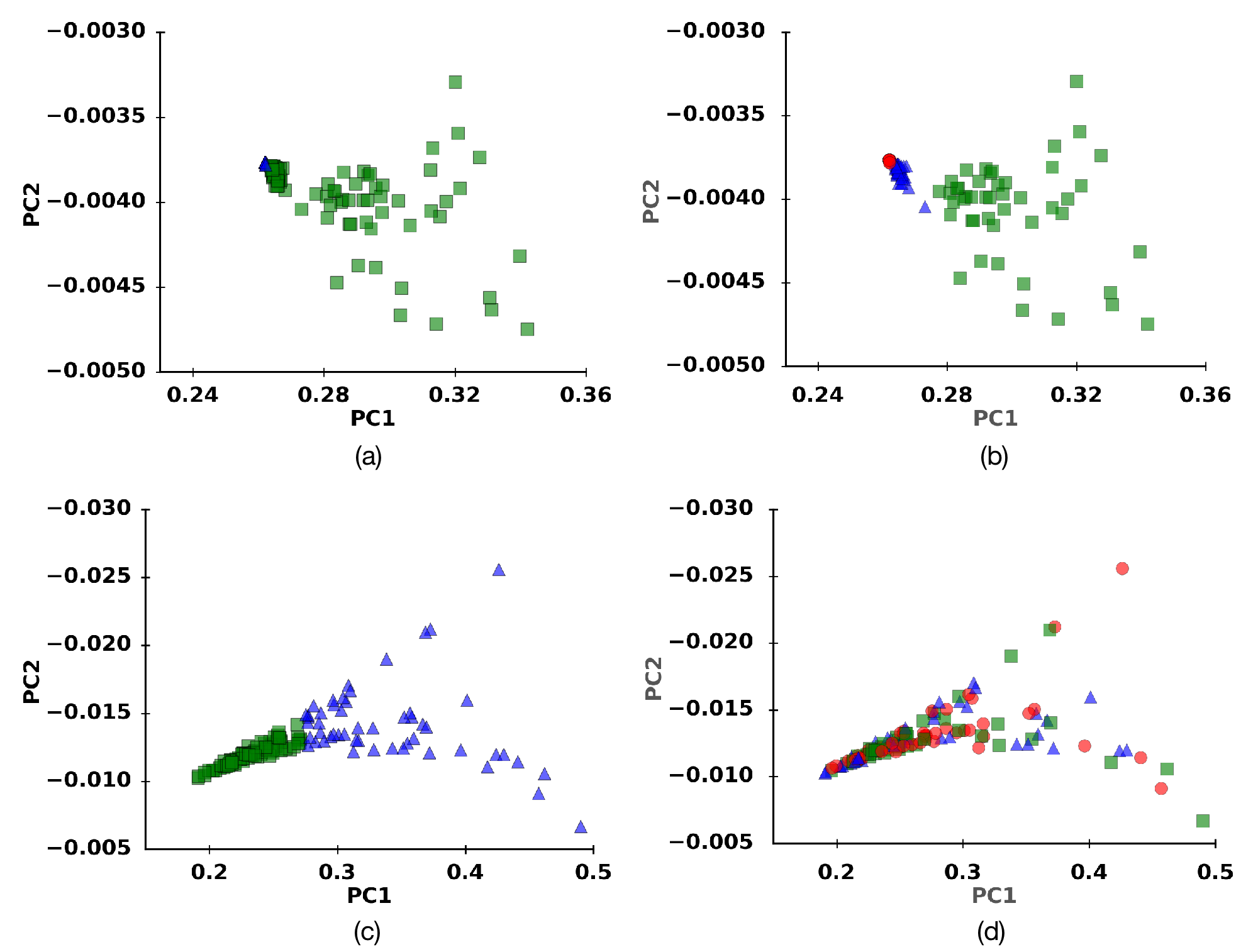}
\caption{ {\textbf{Large-reload {\it vs.} small-reload DIC-testing -- Example of PCA projection results for thin films of $\boldsymbol{w=0.25}~\boldsymbol{\mu \rm{m}}$:}}. $\omega_1$,$\omega_1$ auto-correlations. The colors follow the definition of Fig.~\ref{fig:width2}. (a) Small DIC-test reloaded state strain (0.1 $\%$), with clustering algorithm applied to data set (b) Small DIC-test reloaded state strain (0.1 $\%$), without a clustering algorithm applied to data set. Actual representation of data set. (c) Large DIC-test reloaded state strain (1 $\%$) with a clustering algorithm applied to the data set. (d) Large DIC-test reloaded state (1 $\%$) without a clustering algorithm applied to the data set. Actual representation of the data set.}
\label{fig:inv-noninvcomp2}
\end{figure}

This section contains a direct connection between small and large DIC-testing strain when reloading the samples. In Figs.~\ref{fig:inv-noninvcomp2} (a) we show the results for sample of $w=0.5$ $\micron$ when we use our classifier. In (b) we show how a correct classification should look like for the same samples, when reloading to 0.1 $\%$ DIC-testing strain. In contrast, Figs.~\ref{fig:inv-noninvcomp2} (c) and (d) show samples when reloading to 1 $\%$ DIC-testing strain. While the sample distinction is applied with (c) showing classified data and (d) showing how a correct classification should be, we can also observe that due to the high reload strain value, we added plastic memory to the samples. Thus, we completely changed the PCA projections of our samples.
  
\newpage
\subsection{Dependence of unsupervised learning capacity on pre-processing aspects.}

\begin{figure}[H] \centering
\includegraphics[width=0.68\textwidth]{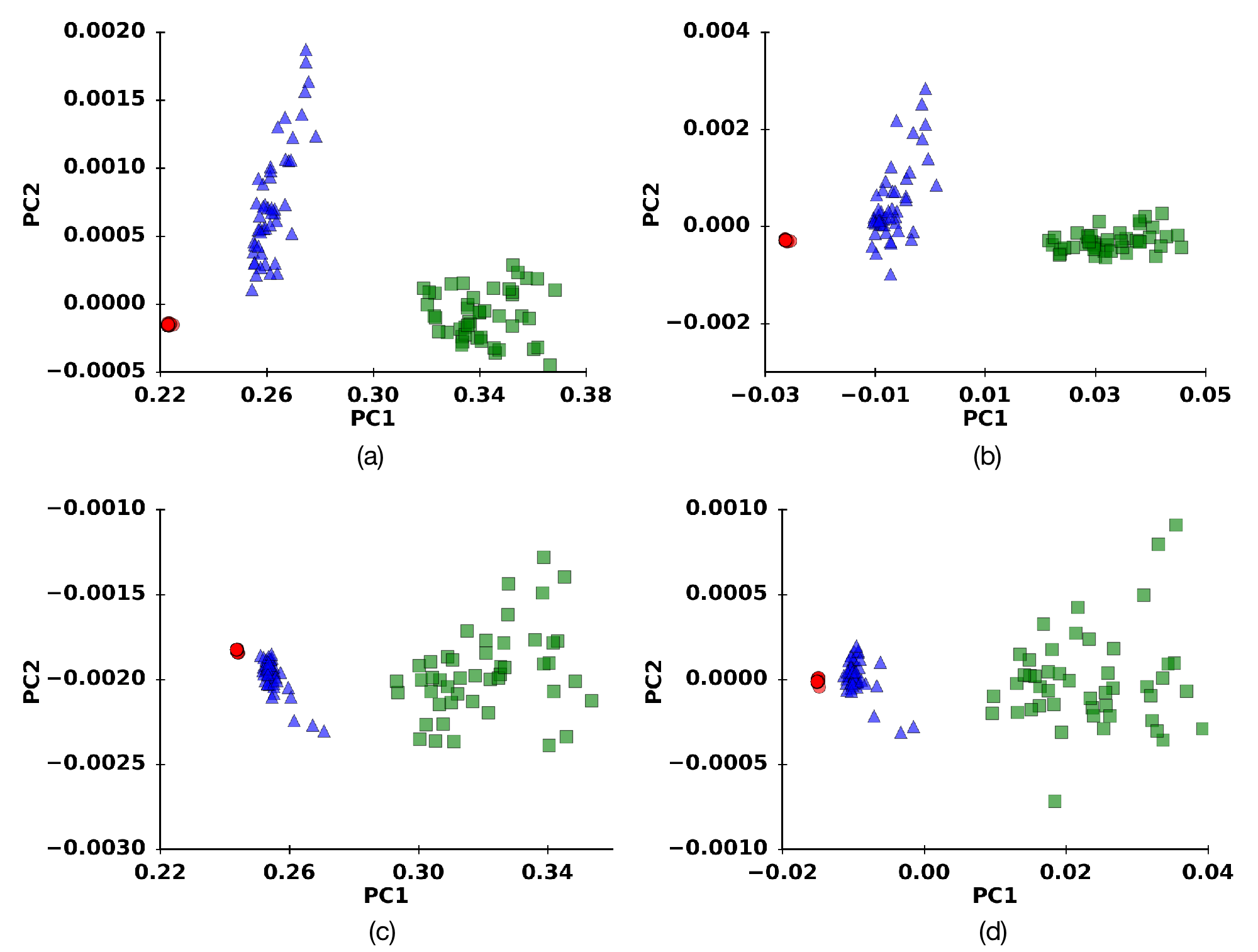}
\caption{ {\textbf{Auto-correlations {\it vs.} cross-correlations for pre-processing -- More examples of PCA projection maps for $\boldsymbol{w=1, 2}$~$\boldsymbol{\mu \rm{m}}$: }} The colors follow the definition of Fig.~\ref{fig:width2}. (a) $\omega_1$,$\omega_1$ auto-correlations. $w=2 \micron$ (b) $\omega_2$,$\omega_2$ auto-correlations. $w=2 \micron$ (c) $\omega_1$,$\omega_1$ auto-correlations. $w=1 \micron$ (d) $\omega_2$,$\omega_2$ auto-correlations. $w=1 \micron$}
\label{fig:w2w1icorr}
\end{figure}

\begin{figure}[H] \centering
\includegraphics[width=0.72\textwidth]{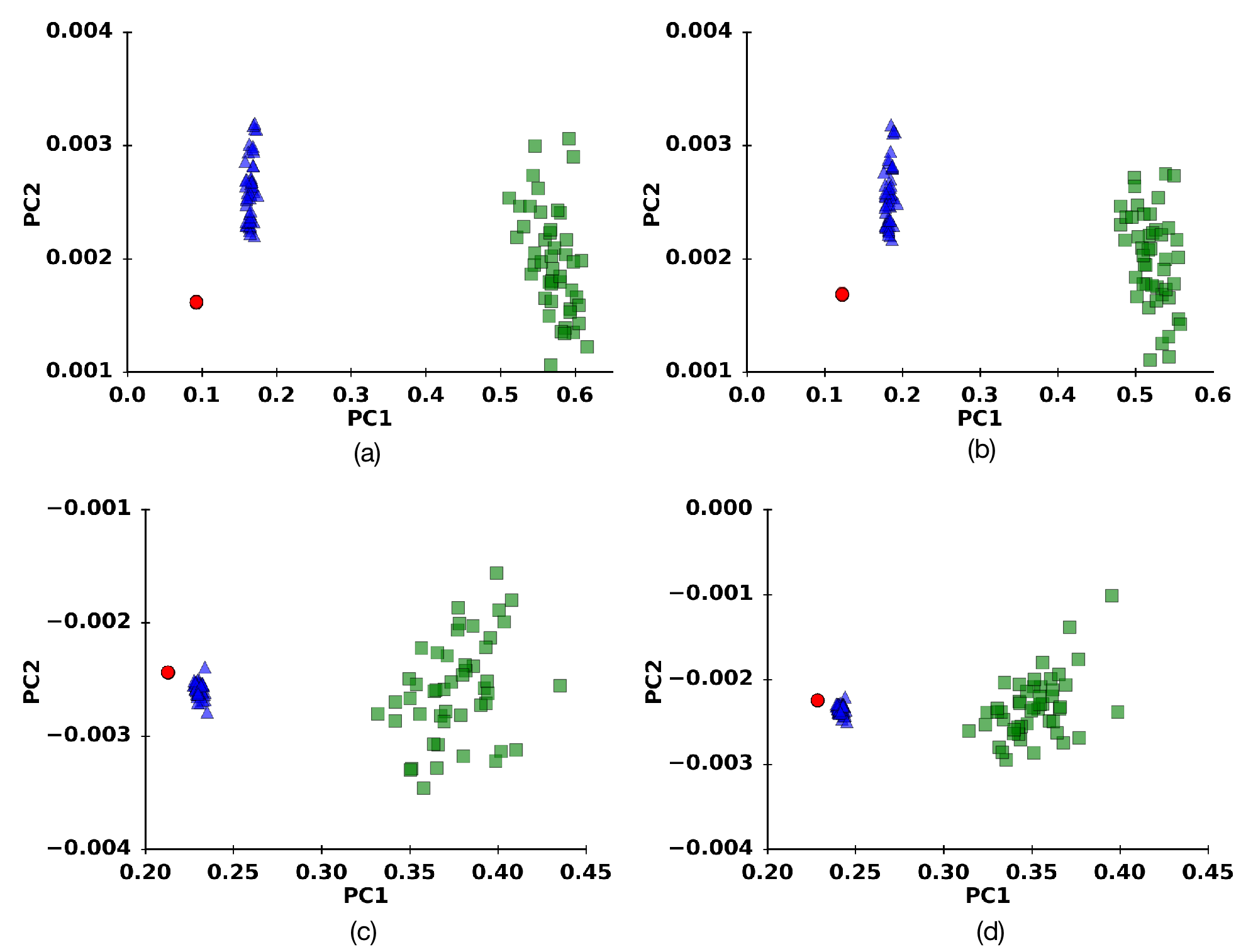}
\caption{ {\textbf{Combined plastic strain calculation and $2^{nd}$ strain-tensor invariant -- Examples of PCA projection maps:}} The colors follow the definition of Fig.~\ref{fig:width2}. (a) $\omega_1$,$\omega_1$ auto-correlations. w=2$ \micron$. Isotropic measurement. (b) $\omega_1$,$\omega_1$ auto-correlations. $w=2 \micron$. $J_2$ invariant. (c) $\omega_1$,$\omega_1$ auto-correlations. $w=1 \micron$. Isotropic measurement. (d) $\omega_1$,$\omega_1$ auto-correlations. $w=1 \micron$. $J_2$ invariant.}
\label{fig:plastj2}
\end{figure}

The final section of this appendix, shows more results for different preprocessing aspects. 
It corresponds to Section 3.8 of the main text. Fig.~\ref{fig:w2w1icorr} shows the classified 
results different types of auto-correlations, for samples of w=1 and w=2 $\micron$. Fig.~\ref{fig:plastj2} shows results when using the plastic strain as the microstructural state variable (see Section 3.8) and applying the correlation statistics through the $\phi$ isotropic measurement for the local strain or the $J_2$ invariant. 

\begin{thebibliography}{100} % 55 (2007) p.4067.

\bibitem[Frazier, 2014]{frazier2014} W. E. Frazier, Metal additive
manufacturing: a review. Journal of Materials Engineering and
Performance, 2014, 23(6), 1917-1928.

\bibitem[Bigoni and Hueckel, 1991]{bigoni1991} D. Bigoni and
T. Hueckel, Uniqueness and localization­ --I. Associative and
non-associative elastoplasticity. Int. J. Solids and Structures, 1991,
28 (2), 197-213.

\bibitem[Pedregosa et~al., 2011]{pedregosa2011} F. Pedregosa, G. Varoquaux, A. Gramfort, V. Michel, B. Thirion, O. Grisel, M. Blondel, P. Prettenhofer, R. Weiss, V. Dubourg, J. Vanderplas, A. Passos, D. Cournapeau, M. Brucher, M. Perrot and E. Duchesnay, Scikit-learn: Machine Learning in Python, J. Mach. Learn. Res., 2011, 12, 2825-2830.

\bibitem[Rasmussen and Nickisch, 2010]{rasmussen} C.E. Rasmussen, H. Nickisch, Gaussian Processes for Machine Learming (GMPL) Toolbox, J. Mach. Learn. Res., 2010, 11, 3011-3015.

\bibitem[Asaro and Lubarda, 2006]{asaro-book} R. Asaro and V. Lubarda,
Mechanics of solids and materials, Cambridge University Press, 2006.

\bibitem[Chen and Boehlert, 2013]{chen2013} Z. Chen and C.J. Boehlert,
Evaluating the Plastic Anisotropy of AZ31 Using Microscopy Techniques,
2013, J. Mater., 65 (9), 1237-1244, doi: 10.1007/s11837-013-0672-6.

\bibitem[Chen et~al., 2012]{chen2012} Z. Chen, A. Shyam, J. Huang,
R.F. Decker, S.E. LeBeau and C.J.  Boehlert, The Small Fatigue Crack
Growth Behavior of an AM60 Magnesium Alloy, Metall. Mater. Trans. A,
2013, 44 (2), 1045-1058, doi: 10.1007/s11661-012-1449-1.

\bibitem[Jon Shlens, 2003]{shlens2003} J. Shlens, A Tutorial on Principal Component Analysis Derivation, Discussion and Singular Value Decomposition, 2003

\bibitem[Chen et~al., 2011]{chen2011} Z. Chen, J. Huang, R.F. Decker,
S.E. LeBeau, L.R. Walker, O.B. Cavin, T.R. Watkins and C.J. Boehlert,
The Effect of Thermomechanical Processing on the Tensile, Fatigue, and
Creep Behavior of Magnesium Alloy AM60, Metall. Mater. Trans. A, 2011,
42 (5), 1386-1399.

\bibitem[Khademi et~al., 2015]{khademi2015} V. Khademi, M. Ikeda,
C.J. Boehlert, The Effect of Temperature on the Tensile Deformation
Behavior of a Beta Titanium Allow, an In-Situ SEM Study, Proceedings
of the 13th World Conference on Titanium, John Wiley \& Sons, Inc.,
2015, 1109-1116, doi: 10.1002/9781119296126.ch188

\bibitem[Khademi et~al., 2017]{khademi2017} V. Khademi, C.J. Boehlert,
R.T. Bieler, M. Ikeda, S. Daly, Z. Chen, A correlation between Digital
Image Correlation and Grain Misorientation Distribution Mapping to
capture the localized plastic deformation in a polycrystalline
titanium alloy, TMS Annual Meeting \& Exhibition, San Diego, CA, USA,
2017.

\bibitem[Khademi et~al., 2016]{khademi2016a} V. Khademi, R.T. Bieler,
M. Ikeda, C.J. Boehlert, Influence of Grain Size and Crystallographic
Orientation on Localized Plastic Strain Distribution in
Polycrystalline Beta Titanium Alloys, Materials Science \& Technology
16, Salt Lake City, UT, USA, 2016

\bibitem[Khademi Bieler et~al., 2016]{khademi2016b} V. Khademi,
R.T. Bieler, M. Ikeda, C.J. Boehlert, Quantifying the Local and Global
Misorientation Distributions as a Function of Crystallographic
Orientation and Level of Plastic Strain in Polycrystalline Materials
by Utilizing EBSD Mapping, TMS 16, Nashville, TN, USA, 2016

\bibitem[Z. Chen, 2012]{chendiss} Z. Chen, A Study of the Effects of
Processing and Alloying on the Microstructure and Deformation Behavior
of Wrought Magnesium Alloys, Thesis (Ph.D., Materials Science
Engineering) - Michigan State University, 2012.

\bibitem[Shan et~al., 2008]{shan2008} Z.W. Shan, R.K. Mishra, S.S
Asif, O.L. Warren and A.M. Minor, Mechanical annealing and
source-limited deformation in submicrometre-diameter Ni
crystals. Nature materials, 2008, 7 (2), 115-119.

\bibitem[Carroll et~al., 2013]{carroll2013} J. D. Carroll, W. Abuzaid,
J. Lambros and H. Sehitoglu, High resolution digital image correlation
measurements of strain accumulation in fatigue crack growth,
Int. J. Fatigue, 2013, 57, 140-150.

\bibitem[Kalidindi, 2015]{kalidindi2015} S.R. Kalidindi,J.A. Gomberg, Z.T. Trautt and C.A. Becker, Application of data science tools to quantify and distinguish between structures and models in molecular dynamics datasets, Nanotechnology, 2015, 26 (34), 344006, doi: 10.1088/0957-4484/26/34/344006

\bibitem[Kalidindi, 2012]{kalidindi2012} S.R. Kalidindi,
Computationally-Efficient Fully-Coupled Multi-Scale Modeling of
Materials Phenomena Using Calibrated Localization Linkages, ISRN
Materials Science, 2012, Article ID 305692.

\bibitem[Fast and Kalidindi, 2011]{fastkalidindi} T.N. Fast and
S.R. Kalidindi, Formulation and Calibration of Higher-Order Elastic
Localization Relationships Using the MKS Approach, Acta Materialia,
2011, 59 (11), 4595-4605.

\bibitem[Fast, 2011]{fast2011} T.N. Fast, Developing higher-order
materials knowledge systems, Thesis (PhD, Materials engineering) –
Drexel University, 2011, doi:1860/4057.

\bibitem[Wheeler et~al., 2014]{wheeler2014} D. Wheeler , D. Brough ,
T. Fast, S.R. Kalidindi and A. Reid, PyMKS: Materials Knowledge System
in Python, 2014.

\bibitem[Mattis and Swendsen, 2008]{mattis2008} D.C. Mattis and R. Swendsen, Statistical Mechanics Made Simple, 2nd Edition, World Scientific Publishing Co. Pte. Ltd., 2008

\bibitem[Niezqoda et~al., 2008]{niezqoda2008} S.R. Niezgoda,
D.T. Fullwood and S.R. Kalidindi, Delineation of the Space of 2-Point
Correlations in a Composite Material System, Acta Materialia, 2008, 56
(18), 5285–5292.

\bibitem[Niezqoda et~al., 2010]{niezqoda2010} S.R. Niezgoda,
D.M. Turner, D.T. Fullwood and S.R. Kalidindi, Optimized Structure
Based Representative Volume Element Sets Reflecting the
Ensemble-Averaged 2-Point Statistics, Acta Materialia, 2010, 58 (13),
4432–4445.

\bibitem[Russell and Norvig, 1995]{russell1995} S. Russell and
P. Norvig, Artificial Intelligence: A Modern Approach (2nd ed.) 1995,
Prentice Hall. ISBN 978-01379039

\bibitem[Werning et~al., 2010]{werning2010} M.N. Wernick, Y. Yang,
J.G. Brankov, G. Yourganov and S.C. Strother, Machine Learning in
Medical Imaging, IEEE Signal Processing Mag., 2010, 27 (4), pp. 25–38.

\bibitem[James et~al., 2013]{james2013} G. James, D. Witten, T. Hastie
and R. Tibshirani, An Introduction to Statistical Learning, 2013, New
York: Springer
 
\bibitem[Friedman, 1998]{friedman1998} J.H. Friedman, Data Mining and
Statistics: What's the connection?, 29th Symposium on the interface,
1998.

\bibitem[Wang et~al., 2006]{ghoniem2006} Z.Q. Wang, N.M. Ghoniem,
S.Swaminarayan and R. LeSar, A parallel algorithm for 3D dislocation
dynamics, J. Comput. Phys., 2006, 219 (2), 608–621.

\bibitem[Asaro and Lubarda, 2006]{asarolubarda} R. Asaro and
V. Lubarda, Mechanics of Solids and Materials, Cambridge: Cambridge
University Press, 2006.

\bibitem[Deshpande et~al., 2012]{deshpande2012} C. Ayas,
V.S. Deshpande, and M. Geers, Tensile response of passivated films
with climb-assisted dislocation glide, J. Mech. Phys. Solids, 2012,
60, 1626-1643.

\bibitem[Deshpande et~al., 2013]{deshpande2013} K. Danas and
V.S. Deshpande, Plane-strain discrete dislocation plasticity with
climb-assisted glide motion of dislocations,
Model. Simul. Mater. Sci. Eng., 2013, 21(4).

\bibitem[Vlassak et~al., 2014]{vlassak2014} K.M Davoudi, L. Nicola and
J.J. Vlassak, Bauschinger effect in thin metal films: Discrete
dislocation dynamics study, J. Applied Phys., 2014, 115 (1).

\bibitem[Nicola et~al., 2006]{nicola2006} L. Nicola, Y.Xiang,
J.J. Vlassak, E. Van der Giessen and A. Needleman, Plastic deformation
of freestanding thin films: experiments and modeling,
J. Mech. Phys. Solids, 2006, 54 (10), 2089-2110.

\bibitem[Shishvan and Van der Giessen, 2010]{shishvan2010}
S.S. Shishvan, S. Mohammadi, M. Rahimian and E. Van der Giessen,
Plane-strain discrete dislocation plasticity incorporating anisotropic
elasticity, Int. J. Solids and Structures, 2011, 48 (2), 374-387.


\bibitem[Van der Giessen et~al., 2006]{vandergiessen2006}
D.S. Balint,V.S Deshpande, A. Needleman and E. Van der Giessen, Size
effects in uniaxial deformation of single and polycrystals: a discrete
dislocation plasticity analysis, Model. Simul. Mater. Sci. Eng., 2006,
14, 409-422.

\bibitem[Fullwood et~al., 2009]{fullwood2009} D.T. Fullwood,
S.R. Kalidindi, and B.L. Adams, Second - Order Microstructure
Sensitive Design Using 2-Point Spatial Correlations, Chapter 12 in
Electron Backscatter Diffraction in Materials Science, 2nd Edition ,
Eds. A. Schwartz, M. Kumar, B. Adams, D. Field, Springer, NY, 2009.

\bibitem[Mitchell, 2006]{mitchell2006}Mitchell T., The discipline of
machine learning, Carnegie Mellon University, 2006, (Technical Report
CMUML-06-108).

\bibitem[Papanikolaou et~al., 2017]{papanikolaou2017} S. Papanikolaou,
H. Song and E. Van der Giessen, Obstacles and sources in dislocation
dynamics: Strengthening and statistics of abrupt plastic events in
nanopillar compression, J. Mech. Phys. Solids, 2017, 102, 17-29.

\bibitem[Hirth and Lothe, 1982]{hirth1982} J.P. Hirth and J. Lothe,
Theory of dislocations, 1982, 2nd ed. (John Wiley and Sons, New York).

\bibitem[Van der Giessen and Needleman, 1995]{vandergiessen1995}
E. Van der Giessen and A. Needleman, Discrete dislocation plasticity:
a simple planar model, Model. Simul. Mater. Sci. Eng., 1995, 3 (5),
689.

\bibitem[Mueller et~al., 2016]{mueller2016} T. Mueller, A.G. Kusne,
and R. Ramprasad, Machine Learning in Materials Science, in Reviews in
Computational Chemistry, Volume 29 (eds A. L. Parrill and
K. B. Lipkowitz), John Wiley $\&$ Sons, Inc, Hoboken, NJ.

\bibitem[Anuta, 1970]{anuta1970} P.E Anuta, Spatial registration of
multispectral and multitemporal digital imagery using fast Fourier
transform techniques, IEEE Trans. Geosci. Electron., 1970, GE-8,
353–368.

\bibitem[Keating et~al., 1975]{keating1975} T.J. Keating , P.R. Wolf
and F.L. Scarpace , An Improved Method of Digital Image Correlation,
Photogrammetric Engineering and Remote Sensing, 1975, 41(8), 993–1002.

\bibitem[McCormick and Lord, 2012]{mccormick2012} N. McCormick and
J.D. Lord, Digital Image Correlation, Materials Today, 2012, 13(12),
52-54.

\bibitem[Roux et~al., 2009]{roux2009} S. Roux, J. Réthoré and F. Hild,
Digital Image Correlation and Fracture: An Advanced Technique for
Estimating Stress, J. Phys. D: Applied Physics, 2009, 42(21).

\bibitem[Berry and Sauer, 2017]{berry2017} T. Berry and T. Sauer,
Consistent Manifold Representation for Topological Data Analysis,
2017.

\bibitem[Makarov et~al., 2001]{makarov2011} P.V. Makarov,
S. Schmauder, O.I. Cherepanov, I.Yu. Smolin, V.A. Romanova,
R.R. Balokhonov, D.Yu. Saraev, E. Soppa, P. Kizler, G. Fischer, S. Hu,
M. Ludwig, Simulation of elastic-plastic deformation and fracture of
materials at micro-, meso- and macrolevels, Theoretical and Applied
Fracture Mechanics, 2001, 37(1-3), 183-244.

\bibitem[Yates and Ribeiro-Neto, 2011]{yates2011} R. Baeza-Yates and B. Ribeiro-Neto, 2011, Modern Information Retrieval, Addison Wesley, pp. 327-328.

\bibitem[David Powers, 2011]{powers2011} D.M.W. Powers, Evaluation: From Precision, Recall and F-Measure to ROC, Informedness, Markedness $\&$ Correlation, J. Mach. Learn. Tech., 2011, 2(1), 37â 63.

\bibitem[Press, 2007]{nr} W.H. Press, Numerical recipes 3rd edition: The art of scientific computing, Cambridge university press, 2007.

\end{thebibliography}
\end{document}